\documentclass[superscriptaddress,amssymb,prx,10pt,twocolumn]{revtex4-1}
\usepackage{amsmath}    
\usepackage{graphicx}   
\usepackage{verbatim}   
\usepackage{color}      
\usepackage{subfigure}
\usepackage{gensymb}
\usepackage{textcomp}
\usepackage[bookmarks=false]{hyperref}   
\usepackage{longtable}
\usepackage{bm}
\usepackage{color}
\usepackage{epstopdf}
\usepackage{physics} 
\usepackage{natbib} 

\hypersetup{
  colorlinks   = true, 
  urlcolor     = blue, 
  linkcolor    = blue, 
  citecolor   = blue 
}

\newcommand{\YBCO}{\ensuremath{\mathrm{YBa_2Cu_3O_7}}}
\newcommand{\LCMO}{\ensuremath{\mathrm{La_{2/3}Ca_{1/3}MnO_3}}}

\newcommand{\LSAT}{\ensuremath{\mathrm{Sr_{0.7}La_{0.3}Al_{0.65}Ta_{0.35}O_3}}}
\newcommand{\LAO}{\ensuremath{\mathrm{LaAlO_3}}}
\newcommand{\STO}{\ensuremath{\mathrm{SrTiO_3}}}
\newcommand{\YLa}{\ensuremath{\mathrm{YL\_1}}}
\newcommand{\YLb}{\ensuremath{\mathrm{YL\_2}}}
\newcommand{\YLc}{\ensuremath{\mathrm{YL\_3}}}
\newcommand{\mB}{\ensuremath{\mu_\mathrm{B}}}

\newcommand{\tcurie}{$T^{Curie}$}
\newcommand{\vi}{\mathbf{i}}
\newcommand{\vj}{\mathbf{j}}

\begin{document}
\title{\textbf{X-ray absorption study of the ferromagnetic Cu moment at the $\mathbf{{YBa_2Cu_3O_7}/{La_{2/3}Ca_{1/3}MnO_3}}$ interface and the variation of its exchange interaction with the Mn moment}} 

\author{K. Sen}
\email{kaushik.sen@unifr.ch}
\author{E. Perret}
\affiliation{University of Fribourg, Department of Physics and Fribourg Center for Nanomaterials, Chemin du Mus\'ee 3, CH-1700, Fribourg, Switzerland}
\author{A. Alberca}
\affiliation{University of Fribourg, Department of Physics and Fribourg Center for Nanomaterials, Chemin du Mus\'ee 3, CH-1700, Fribourg, Switzerland}
\affiliation{Swiss Light Source, Paul Scherrer Institut, CH-5232, Villigen PSI, Switzerland}
\author{M.A. Uribe-Laverde}
\affiliation{University of Fribourg, Department of Physics and Fribourg Center for Nanomaterials, Chemin du Mus\'ee 3, CH-1700, Fribourg, Switzerland}
\author{I. Marozau}
\affiliation{University of Fribourg, Department of Physics and Fribourg Center for Nanomaterials, Chemin du Mus\'ee 3, CH-1700, Fribourg, Switzerland}
\author{M. Yazdi-Rizi}
\affiliation{University of Fribourg, Department of Physics and Fribourg Center for Nanomaterials, Chemin du Mus\'ee 3, CH-1700, Fribourg, Switzerland}
\author{B.P.P. Mallett}
\affiliation{University of Fribourg, Department of Physics and Fribourg Center for Nanomaterials, Chemin du Mus\'ee 3, CH-1700, Fribourg, Switzerland}
\author{P. Marsik}
\affiliation{University of Fribourg, Department of Physics and Fribourg Center for Nanomaterials, Chemin du Mus\'ee 3, CH-1700, Fribourg, Switzerland}
\author{C. Piamonteze}
\affiliation{Swiss Light Source, Paul Scherrer Institut, CH-5232, Villigen PSI, Switzerland}
\author{Y. Khaydukov}
\affiliation{Max-Planck-Institut f\"{u}r Festk\"orperforschung, Stuttgart, 70569, Germany}
\affiliation{Max Planck Society Outstation at the Forschungsneutronequelle Heinz Maier-Leibnitz (FRM-II), D-85747 Garching, Germany}
\author{M. D\"obeli}
\affiliation{Ion Beam Physics, ETH Zurich, CH-8093, Zurich, Switzerland}
\author{T. Keller}
\affiliation{Max-Planck-Institut f\"{u}r Festk\"orperforschung, Stuttgart, 70569, Germany}
\affiliation{Max Planck Society Outstation at the Forschungsneutronequelle Heinz Maier-Leibnitz (FRM-II), D-85747 Garching, Germany}
\author{N. Bi\v{s}kup}
\affiliation{Departamento de F\'{i}sica Aplicada III, Instituto Pluridisciplinar, Universidad Complutense de Madrid, Spain}
\affiliation{Centro Nacional de Microscop\'{i}a Electr\'{o}nica, Universidad Complutense de Madrid, Spain}
\author{M. Varela}
\affiliation{Departamento de F\'{i}sica Aplicada III, Instituto Pluridisciplinar, Universidad Complutense de Madrid, Spain}
\author{J. Va\v{s}\'atko}
\author{D. Munzar}
\affiliation{Department of Condensed Matter Physics, Faculty of Science, and Central European Institute of Technology, Masaryk University, Kotl\'a\v{r}sk\'a 2, 61137 Brno, Czech Republic}
\author{C. Bernhard}
\email{christian.bernhard@unifr.ch}
\affiliation{University of Fribourg, Department of Physics and Fribourg Center for Nanomaterials, Chemin du Mus\'ee 3, CH-1700, Fribourg, Switzerland}


\begin{abstract}
With x-ray absorption spectroscopy and polarized neutron reflectometry we studied how the magnetic proximity effect at the interface between the cuprate high-$T_C$  superconductor {\YBCO} (YBCO) and the ferromagnet {\LCMO} (LCMO) is related to the electronic and magnetic properties of the LCMO layers. In particular, we explored how the magnitude of the ferromagnetic Cu moment on the YBCO side depends on the strength of the antiferromagnetic (AF) exchange coupling with the Mn moment on the LCMO side. We found that the Cu moment remains sizeable if the AF coupling with the Mn moments is strongly reduced or even entirely suppressed. The ferromagnetic order of the Cu moments thus seems to be intrinsic to the interfacial CuO$_2$ planes and related to a weakly ferromagnetic intra-planar exchange interaction. The latter is discussed in terms of the partial occupation of the Cu $3d_{3z^2-r^2}$ orbitals, which occurs in the context of the so-called orbital reconstruction of the interfacial Cu ions.
\end{abstract}
\maketitle

\section{Introduction}
The physical properties of interfaces between complex oxides are of great current interest~\cite{Hwang2012}. A prominent example is the interface between the two band insulators {\LAO} and {\STO} at which highly mobile carriers are confined and give rise to electronic and superconducting phenomena that can be tuned with a gate voltage~\cite{Thiel2006,Reyren2007}. Another interesting example involves the magnetic proximity effect (MPE) at the interface between the cuprate high-$T_C$ superconductor (SC) {\YBCO} (YBCO) and ferromagnet (FM) {\LCMO} (LCMO)~\cite{Sefrioui2003,Przyslupski2004,Pena2005,Stahn2005, Hoffman2005,Chakhalian2006, Hoppler2009, Satapathy2012, Giblin2012, Visani2012, Golod2013, Kalcheim2014}. With polarized neutron reflectometry (PNR) it was found that, in the vicinity of the interface, the FM order of the Mn moments is strongly suppressed~\cite{Stahn2005,Hoffman2005,Satapathy2012,Uribe2013}. This phenomenon has been discussed in terms of a \textit{\lq{dead layer}\rq} or a \textit{\lq{depleted layer}\rq}. In addition, it was shown with x-ray magnetic circular dichroism (XMCD) that the interfacial Cu ions acquire a ferromagnetic moment of about 0.2\,{\mB} which is antiparallel to the one of Mn~\cite{Chakhalian2006,Werner2010,Satapathy2012,Uribe2014}. Recent x-ray resonant magnetic reflectometry (XRMR) studies have demonstrated that these Cu moments reside in the interfacial CuO$_2$ planes~\cite{Alberca2015}. Furthermore, x-ray linear dichroism (XLD) studies revealed that the interfacial Cu ions undergo an orbital reconstruction which yields a large hole density in the Cu $3d_{3z^2-r^2}$ orbitals (that are almost fully occupied in the bulk)~\cite{Chakhalian2007,Uribe2014}. Both the orbital reconstruction and the magnetic moment of the interfacial Cu ions have been explained in terms of a strong hybridization between the Cu and Mn $3d_{3z^2-r^2}$ orbitals which leads to an antiferromagnetic exchange interaction (AF-EI) between the Cu and Mn spins~\cite{Chakhalian2007}. In this context, the ferromagnetic Cu moment is induced by the AF-EI with the Mn moments, and one expects that the magnitude of this Cu moment is proportional to the strength of this coupling~\cite{Chakhalian2006,Salafranca2010,Salafranca2014}. {\par}
We have performed XMCD, XLD and PNR measurements on a series of YBCO/LCMO multilayers (MLs) for which the strength of the AF-EI between Cu and Mn has been altered by changing the electronic and magnetic properties of the LCMO layers. As described in the experimental section, this goal has been achieved by changing the growth and annealing conditions, as well as the thickness of the LCMO layers. To our surprise, we have found that the magnitude of the FM Cu moment is almost independent of the strength of the AF-EI with Mn, i.e., it remains sizeable if the AF-EI is strongly reduced or even absent. This suggests that the FM order of the Cu moments is not induced by the coupling to the FM Mn moments of LCMO or by a transfer of spin-polarized charge carriers. Instead, this Cu moment seems to be intrinsic to the interfacial CuO$_2$ planes. This circumstance is discussed in terms of a FM intra-planar Cu-Cu exchange interaction that is brought about by the partially occupied Cu $3d_{3z^2-r^2}$ orbitals.
\section{Experiment}
\subsection{Growth and annealing}{\label{pld}}
The YBCO/LCMO (YL) MLs have been grown on $(001)$-oriented {\LSAT} (LSAT) substrates using the pulsed laser deposition (PLD) technique. The \textit{layer-by-layer} growth mode and the overall layer thickness have been monitored with \textit{in-situ} reflection high energy electron diffraction (RHEED) as described in Ref.~\cite{Malik2012}. Three different kinds of MLs, in the following we denote them as {\YLa}, {\YLb} and {\YLc}, have been prepared for which the LCMO layers have substantially different electronic and magnetic properties. This has been achieved by changing the growth and annealing conditions as to alter the concentration of oxygen and cation vacancies of the LCMO layers. {\par}
For all samples, we preheated the LSAT substrates to 825\,$^{\circ}$C in $0.34$\,mbar of O$_2$ for 30~minutes prior to the depositions in order to cure the mechanically polished surface. {\par}
The samples of type {\YLa} have been grown with 10 bilayer (BL) repetitions following a similar procedure as described in Ref.~\cite{Malik2012}. The YBCO layers with a thickness of $d\approx$10\,nm and the LCMO layers with $d\approx$10\,nm were grown with an oxygen partial pressure of 0.34\,mbar, a laser fluence of $2.4$\,J/cm$^{2}$, and a repetition rate of $7$\,Hz. After deposition, they were cooled to 700\,$^{\circ}$C at a rate of 10\,$^{\circ}$C/min while the oxygen partial pressure was gradually increased to $1$\,bar. Subsequently, the samples were further cooled to 485\,$^{\circ}$C at a rate of 30\,$^{\circ}$C/min where they were kept for one hour. The temperature was then slowly decreased to room temperature before removing them from the PLD chamber. To ensure a full oxygenation of the YBCO layers, we performed an \textit{ex-situ} annealing at 485\,$^{\circ}$C in a continuous flow of O$_2$ for 12 hours. {\par}
The samples of type {\YLb} have nominally the same YBCO and LCMO layer thicknesses as {\YLa}, and have been grown with 1, 6 and 10 BL repetitions. These samples have been protected with a capping layer of about 1.5\,nm {\LAO} (LAO). Different O$_2$ partial pressures of 0.34\,mbar and 0.12\,mbar have been used during the growth of the YBCO and LCMO layers, respectively. The laser fluence was kept at 2.0\,J/cm$^2$ and the repetition rate at 2\,Hz. The following \textit{in-situ} cooling and annealing procedures were the same as for {\YLa}, except for a lower cooling rate of only 10\,$^{\circ}$C/min (instead of 30\,$^{\circ}$C/min for {\YLa}) to 485\,$^{\circ}$C. No post annealing was performed since the YBCO layers grown at this lower laser repetition rate have already rather high $T_C$ values (see subsection~\ref{rt-mr-mt-mh}). {\par}
Sample {\YLc} is a single BL with about 19\,nm of YBCO and 5\,nm of LCMO. It is also protected with a LAO capping layer. It was grown with the same PLD parameters as {\YLb}, the only difference concerns the \textit{in-situ} annealing procedure for which the cooling from 700\,$^{\circ}$C to 485\,$^{\circ}$C was done at a faster rate of 30\,$^{\circ}$C/min.
No \textit{ex-situ} annealing was performed to this sample.
\subsection{Structural and chemical characterization}
\begin{figure}[!htbp] 
\centering
\includegraphics{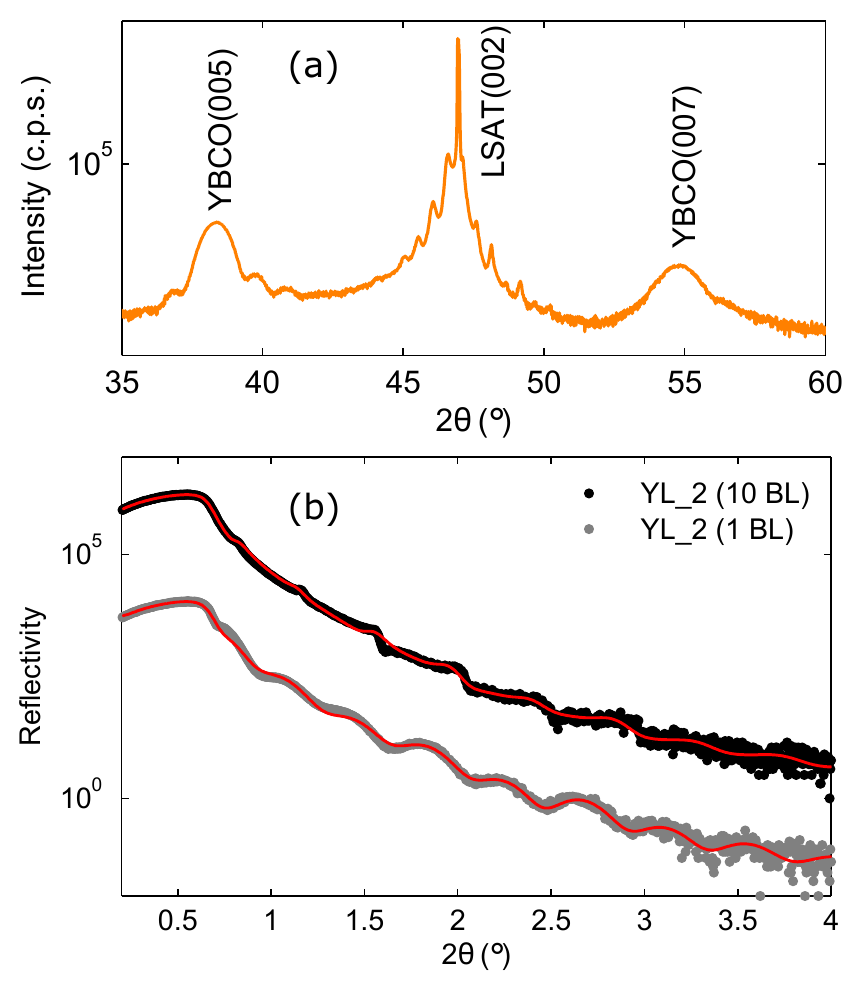}
\caption{\label{fig-xrd} (a) Symmetric $\theta$-$2\theta$ x-ray diffraction curve of {\YLb} (with 10 BL repetitions) along the [00L] direction. Thickness oscillations around the high intensity Bragg peaks testify for the quality of the ML. (b) X-ray reflectivity profiles (symbols) for {\YLb} type samples with 1 and 10 BL repetitions. The solid red lines are the best fits to the data.}
\end{figure}

\begin{figure}[!htbp] 
\centering
\includegraphics{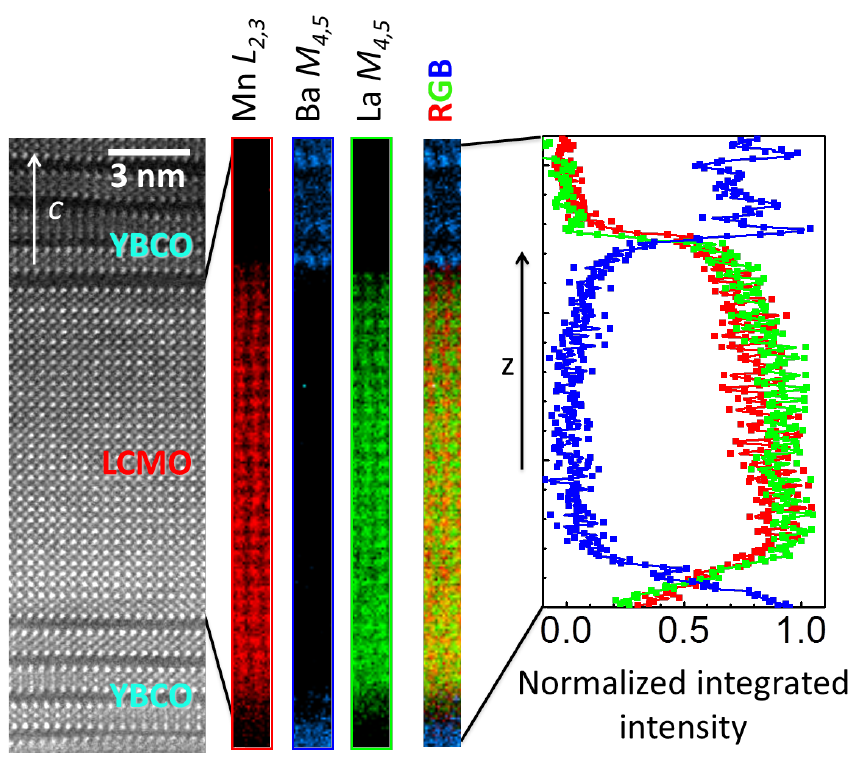}
\caption{\label{fig-tem-eels} Left panel: Atomic resolution, Z-contrast high angle annular dark field image of the YBCO/LCMO/YBCO stacking in a {\YLb} type sample. Right panel:Elemental maps obtained from the analysis of the Mn $L_{2,3}$ (red), Ba $M_{4,5}$ (blue) and La M$_{4,5}$ (green absorption edges). An RGB overlay of the three maps, along with a line profile on a matching color scale of the normalized integrated intensities is also shown. Some spatial drift is visible.}
\end{figure}

{\label{xrd-rbs}}
The structural quality of all the MLs has been confirmed with \textit{in-situ} RHEED and \textit{ex-situ} x-ray diffraction (XRD). The latter also demonstrates the epitaxial growth of the layers with the c-axis of YBCO oriented along the surface normal. Representative XRD patterns for the samples of type {\YLa} can be found in Ref.~\cite{Malik2012} and for {\YLb} in Figure~\ref{fig-xrd}~(a). The fitting of the x-ray reflectivity profiles of the {\YLb} samples with 1 and 10 BL repetitions, shown in Figure~\ref{fig-xrd}~(b), was performed with the GenX software~\cite{Bjorck2007}. It yields a layer thickness of about 9.7\,nm for YBCO and 9.1\,nm for LCMO. The roughness of the YBCO/LCMO interfaces are 6 and 10\,{\AA} for the samples with 1 and 10 BL repetitions, respectively. {\par}
Cross-section high-resolution scanning transmission electron microscopy (STEM) observations of a {\YLb} type sample were carried out in an aberration-corrected JEOL JEM-ARM200cF, operated at 200\,kV and equipped with a cold field emission gun and a Gatan quantum electron energy-loss spectrometer (EELS). The convergence semi-angle was around 35\,mrad, while the collection semi-angle was 28\,mrad, approximately. The specimens were prepared by conventional methods of grinding and Ar-ion milling. Random noise in the EELS data was removed by means of principal component analysis~\cite{Bosman2006}. EELS elemental mapping was performed by integrating the signals under the characteristic elemental edges after background subtraction using a power law. The integration windows were typically around $20$-$30$\,eV wide.  {\par}
STEM-EELS images show coherent, epitaxial interfaces (see Figure~\ref{fig-tem-eels}, left panel). Occasional defects are present such as double CuO chain layers or one unit cell interface steps giving rise to anti phase boundaries, all of these being typical defects observed in YBCO. The CuO chain layers are easily identifiable in the high resolution Z-contrast images due to their slightly darker contrast~\cite{Visani2011,Salafranca2014}. In most cases, the interfaces are found to be symmetric. Regardless of top or bottom relative positions, the dominant atomic plane stacking found is such that a manganite MnO$_2$ plane faces a cuprate BaO plane, as shown by the EELS profiles such as the ones in Figure~\ref{fig-tem-eels}~(right panel). {\par}
Rutherford backscattering (RBS) measurements were performed at the 6\,MV tandem accelerator of the Laboratory of Ion Beam Physics at ETH Zurich using a 2\,MeV $^4$He beam and a silicon PIN diode detector under a backscattering angle of 168$^\circ$~\cite{Doebeli2008}. The experimental data have been analyzed by the RUMP code~\cite{Doolittle1986}. We measured two different types of LCMO films that were grown similar to the LCMO layers in {\YLa} and {\YLb}. To be most suitable for the RBS measurements, the LCMO films were 100\,nm thick and grown on MgO substrates. The obtained stoichiometry of the LCMO samples is listed in Tab.~\ref{rbs-stoi}. Whereas the LCMO layers in {\YLa} are more or less stoichiometric, in {\YLb} there is a significant deficiency of oxygen and a deviation of the cation content from the nominal value. Accordingly, the LCMO layers in {\YLb} are likely to have a somewhat reduced hole content. Note that the uncertainty of the content of the heavier elements, like La, Ca, and Mn, is 1-3\,\%, whereas for oxygen it is up to 5\,\%.
\begin{table}[!htbp]
\centering
\begin{tabular}{l | l | l | l | l |}
\cline{2-5}
 & La & Ca & Mn & O \\
\hline
\multicolumn{1}{|l|}{LCMO in {\YLa}} & $0.66$ & $0.34$ & $0.98$ & $3.05$ \\
\hline
\multicolumn{1}{|l|}{LCMO in {\YLb}} & $0.71$ & $0.29$ & $0.95$ & $2.95$ \\
\hline
\end{tabular}
\caption{Stoichiometry of 100\,nm thick LCMO films grown under similar conditions as {\YLa} and {\YLb}.\label{rbs-stoi}}
\end{table}
\subsection{DC magnetization and electric transport}
\begin{figure}[!htbp] 
\centering
\includegraphics{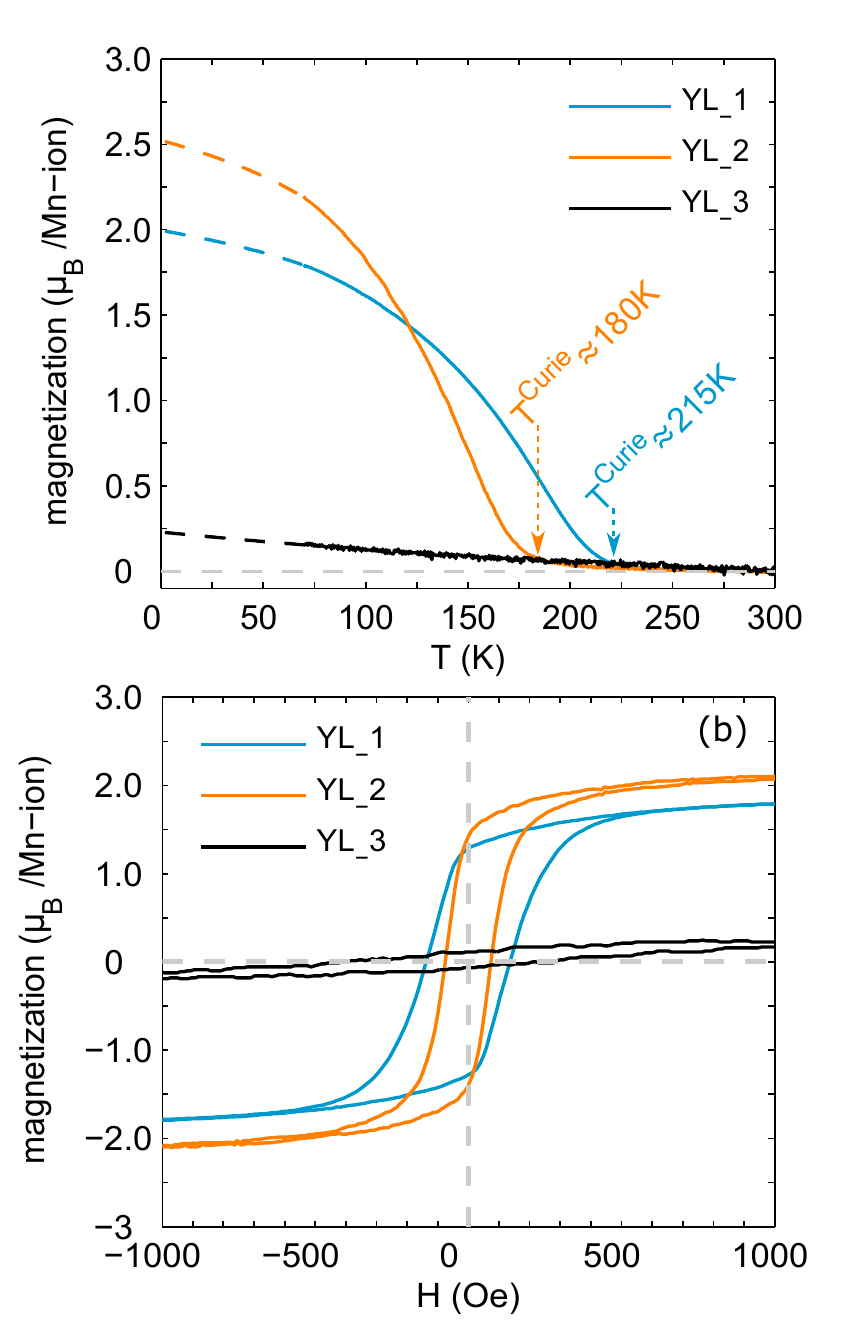}
\caption{\label{fig-mt-vsm} (a) Temperature dependence of the field cooled magnetization at $0.1$\,T for MLs YL\_1-3. The data below $70$\,K have been omitted since they are strongly affected by vortex pinning and related avalances effects that lead to a macroscopic inhomogeneity of the magnetization. Instead we extrapolated the magnetization curves according to $M=M_{S}(\frac{T^{Curie}-T}{T^{Curie}})^{\gamma}$ as shown by the dashed lines. (b) M-H loops for samples YL\_1-3 at $80$\,K measured after field cooling at $9$\,T}
\end{figure}

\begin{figure}[!htbp] 
\centering
\includegraphics{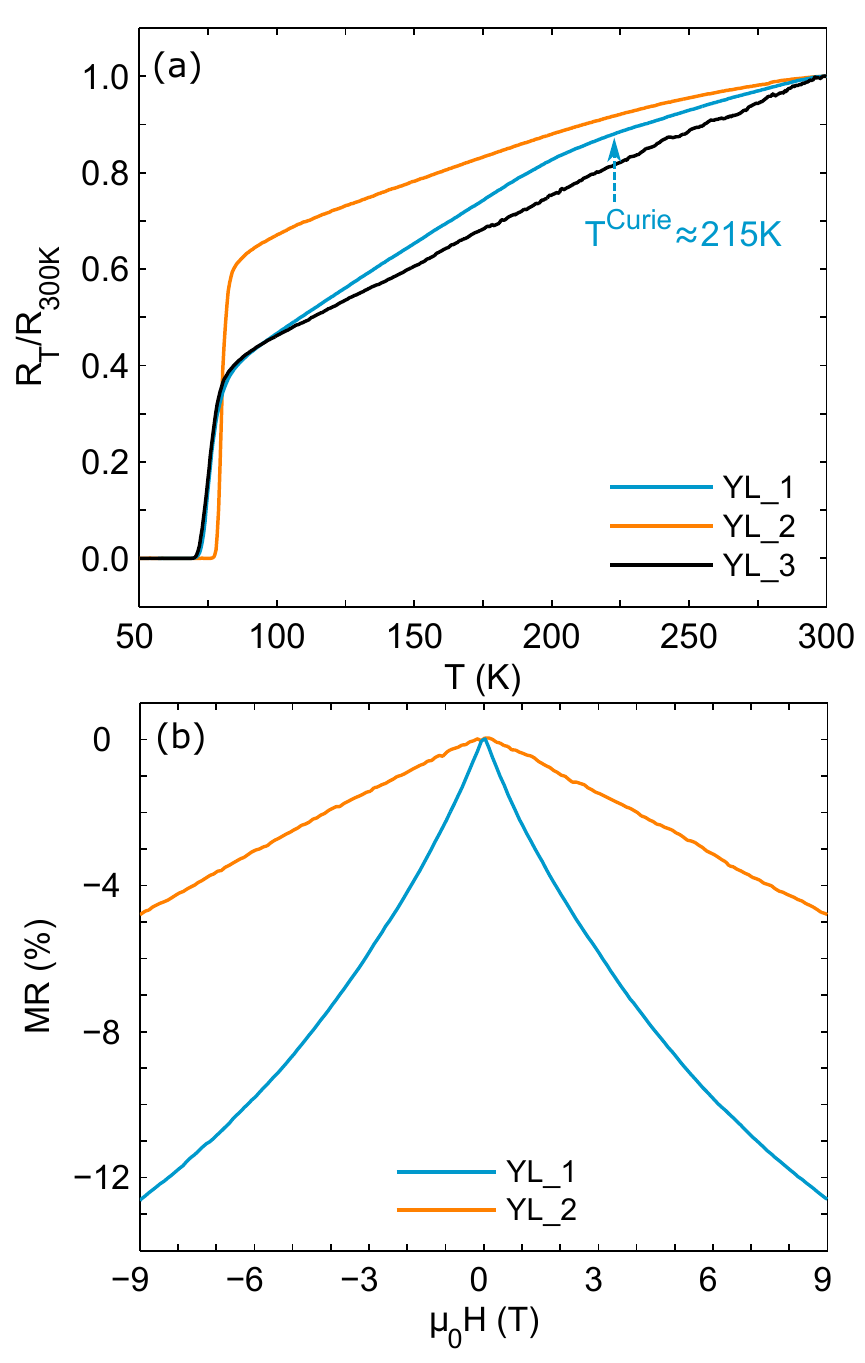}
\caption{\label{fig-rt-rh} (a) Resistance versus temperature in zero magnetic field for YL\_1-3. (b) Magneto-resistance, $\frac{R(H)-R(0)}{R(0)}$ at $150$\,K for {\YLa} and {\YLb}.}
\end{figure}

{\label{rt-mr-mt-mh}}
The DC magnetization has been measured using the Vibrating Sample Magnetometer (VSM) option of a Physical Property Measurement System (PPMS) from Quantum Design. Figure~\ref{fig-mt-vsm} displays the temperature and field dependent DC magnetizations (M-T and M-H curves) of the samples YL\_1-3. The M-T data were acquired during field cooling at a rate of 2\,K/min in 0.1\,T applied parallel to the sample surface. The M-H loops at 80\,K were recorded after cooling the samples in 9\,T. The LCMO layers in {\YLa} have a Curie temperature of {\tcurie}$\approx215$\, K and a sizeable magnetization of 2.0\,{\mB}/Mn. As compared to {\YLa}, the LCMO layers in {\YLb} have a noticeably lower Curie temperature of {\tcurie}$\approx180$\,K and a somewhat higher magnetization of 2.5\,{\mB}/Mn. Due to the very small thickness (5\,nm) of the LCMO layer in {\YLc}, the FM order of the Mn moments is strongly suppressed. The dc magnetization data yield a very small moment of about 0.2\,{\mB}/Mn. To confirm that the suppression of the FM order is due to the reduced LCMO layer thickness, we have grown a BL with a 10\,nm thick LCMO layer under identical conditions and found that it has a FM magnetization comparable to {\YLa} and {\YLb}, and a Curie temperature of {\tcurie}$\approx150$\,K. A similar threshold effect of the FM properties depending on the thickness of the manganite layers has been reported in Ref.~\cite{Monsen2014}. {\par}
The resistance and magneto-resistance were also measured with a PPMS using a four probe method with the wires glued with silver paint to the corners of the sample surface. The current was set to 10\,${\mu}A$ while the voltage was recorded. The temperature was changed at a rate of 2\,K~per~minute, the magnetic field was varied from $-9$\,T to $+9$\,T at a rate of 100\,Oe~per~second. Figure~\ref{fig-rt-rh}~(a) shows that the YBCO layers of YL\_1-3 have fairly sharp superconducting transition with the critical temperatures ($T_C$) of 70-75\,K. On the other hand, the resistance data show that the conductivity of the LCMO layers in {\YLb} is significantly lower than the one in {\YLa}. This is evident from the absence of a kink feature in the R-T curve in Figure~\ref{fig-rt-rh}~(a) and the much lower magneto-resistance effect in the R-H curve in the vicinity of {\tcurie} in Figure~\ref{fig-rt-rh}~(b).
\subsection{X-ray absorption spectroscopy}{\label{xas}}
The x-ray absorption spectroscopy (XAS) measurements have been performed at the XTreme beamline of Swiss Light Source at the Paul Scherrer Institut in Switzerland. The absorption spectra have been recorded simultaneously in the total electron yield (TEY) and the total fluorescence yield (TFY) modes. {\par}
For the XMCD measurements at 2\,K, the samples have been field cooled in 6\,T applied parallel to the incident x-ray beam at a 30$^{\circ}$ incident angle with respect to the sample surface. For each XMCD spectrum, we have changed the polarization of the incoming x-rays (left and right circular) as well as the direction of the applied field, $H_{ext}$. Multiple XMCD measurements have been carried out to check reproducibility and to enhance the signal to noise ratio. Finally, the XMCD is defined as the difference between two absorptions for which the angular momentum of the incoming x-ray photons parallel ($\mu_{+}$) and antiparallel ($\mu_{-}$) to $H_{ext}$. The presented XMCD spectra are normalized with respect to the maximum of $\frac{1}{2}(\mu_{+}+\mu_{-})$, and expressed in percentage. {\par}
The XMCD field scans have been obtained by measuring at two specific energies near the maximum of the $L_3$-edge and off-resonance near the pre-edge, respectively. We have verified that the difference between these two values represents the background subtracted signal. The XMCD field scans are scaled with respect to the corresponding high field XMCD spectra. Due to the remanence of the superconducting magnet, we could not obtain reliable data below ${\pm}0.4$\,T. {\par}
The XLD spectra were obtained at 2\,K and $+0.5$\,T using linearly polarized x-rays with the electric field vector along the vertical ($\sigma$ polarization) and horizontal ($\pi$ polarization) direction with respect to the plane of incidence. The incidence angle correction was performed according to:  $\mu_{ab}=\mu_{\sigma}$; $\mu_c=\sec^2\theta\mu_{\pi}-\tan^2\theta\mu_{\sigma}$; XLD~$=\mu_{ab}-\mu_c$. Multiple sets of measurements were carried out to check reproducibility of the XLD spectra. Finally, the representative XLD spectra have been normalized with respect to the maximum of ${\frac{1}{3}}({2\mu_{ab}+\mu_c})$ and are expressed in percentage.
\subsection{Polarized neutron reflectometry}{\label{pnr}}
The polarized neutron reflectometry (PNR) experiment on the ML {\YLb} with 10 BL repetitions has been performed at the NREX beamline of the FRMII reactor in Munich, Germany, using a monochromatic neutron beam with a wavelength of 4.28\,{\AA}. {\par}
At 300\,K${\textgreater}${\tcurie}$\approx180$\,K, where the sample is not yet ferromagnetic, the reflectivity curve has been measured in unpolarized mode. Subsequently, the sample was cooled to 5\,K in a field 0.1\,T field that was applied parallel to the sample surface. The reflectivity profiles at 5\,K have been obtained for the spin-up and spin-down states of the neutrons. {\par}
The data have been fitted with the same model of block-like nuclear and magnetic potentials as in Refs.~\cite{Satapathy2012,Uribe2013} using the Superfit program from the Max Planck Institute in Stuttgart that is based on a {\lq{Supermatrix formalism}\rq}~\cite{Ruhm1999}. At first, we deduced the structural parameters from the fitting of the 300\,K data. These were later used as input parameters for the fitting of the spin-polarized 5\,K data from which the magnetic depth profile has been obtained.
\subsection{Optical spectroscopy}{\label{opt}}
The optical conductivity of the MLs {\YLa} and {\YLb} has been measured with broad-band spectroscopic ellipsometry. In the near-infrared to ultraviolet range (0.5{-}6.5\,eV), we used a commercial ellipsometer (Woollam VASE) equipped with an ultra-high vacuum liquid He-flow cryostat. In the far-infrared and mid-infrared, we used a home-built setup,~\cite{Bernhard2004} attached to a Bruker 13v FTIR spectrometer with a glowbar source. The correction for the response of the substrate has been performed with the Woollam VASE software~\cite{Woolam}.
\section{Results and Discussions}
In subsection~\ref{suppression_AF-EI} we present the x-ray absorption data (XMCD and XLD) on the MLs YL\_1-3 which reveal that the strength of the antiferromagnetic exchange interaction (AF-EI) between the Cu and Mn moments can be strongly suppressed whereas the magnitude of the Cu moments is hardly reduced.
In subsection~\ref{two-component} we discuss the evidence that the major part of this Cu moment originates from the interfacial Cu ions. In the following, we consider different possibilities to explain the strong reduction of the AF-EI between the interfacial Cu and Mn moments in {\YLb} as compared to {\YLa}.
In subsection~\ref{result-pnr} we show, based on polarized neutron reflectometry data, that this reduction is not related to a corresponding suppression of the ferromagnetic order of the interfacial Mn moments.
In subsection~\ref{result-optics} we present the evidence, based on transport, optical and XLD data, that the strength of the AF-EI is rather related to a change of the electronic properties of the LCMO layers. In particular, that the  orbital polarons in the poorly conducting LCMO layers of {\YLb} are strongly reducing the AF-EI.
Finally, in subsection~\ref{result-yltheory} we provide an explanation of the intrinsic nature of the ferromagnetic Cu moment in terms of the modification of the intra-planar Cu-Cu exchange interaction that is brought about by the so-called orbital reconstruction of the interfacial Cu ions.
\subsection{Suppression of the AF-EI between Cu and Mn} \label{suppression_AF-EI}
\begin{figure*}[!htbp] 
\centering
\includegraphics{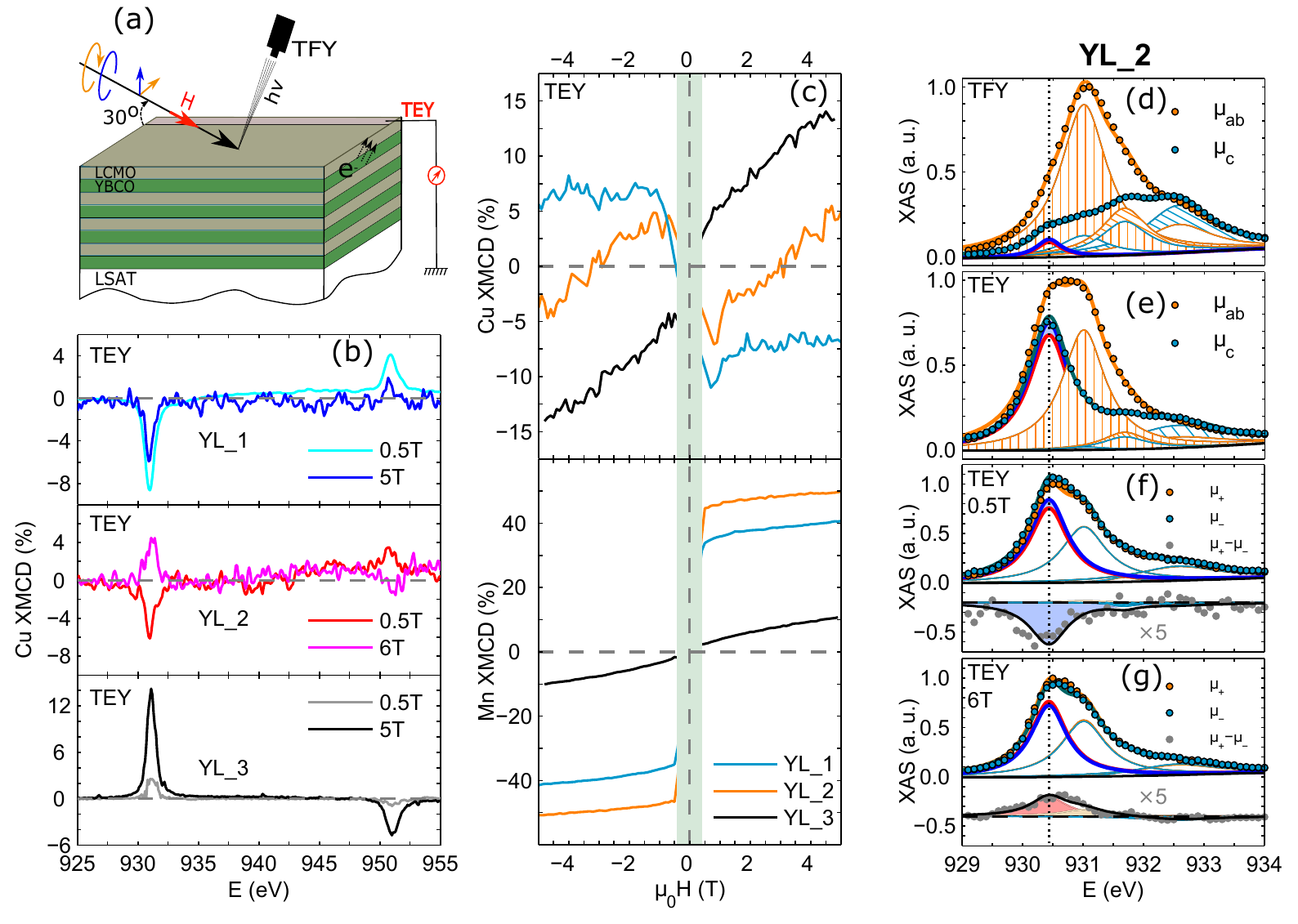}
\caption{\label{Fig-cu-xmcd} (a) Sketch showing the YBCO/LCMO ML and the setup for the XMCD and XLD experiments in TEY and TFY modes. (b) Cu XMCD spectra of the MLs {\YLa}, {\YLb} and {\YLc} measured in TEY mode at low and high magnetic fields. (c) Magnetic field scans of the XMCD at the $L_3$-edges of Cu and Mn in the TEY mode. The shaded area marks the low-field region (below $\pm0.4$\,T) where the remanence of the superconducting magnet inhibits reliable measurements. (d) and (e) Cu XAS curves of {\YLb} with the linear polarization of the x-rays parallel and perpendicular to the CuO$_2$ planes in the TFY and TEY modes, respectively, and a decomposition along the lines of Ref.~\cite{Uribe2014}. (f) and (g) Cu XAS curves of {\YLb} for circular polarizations in the TEY mode and the resulting XMCD curves (lower panel) at $0.5$\,T and $6$\,T, respectively. All data have been taken at $2$\,K. The dotted line along (d)-(g) marks the position of the peak due to the absorption from the interfacial Cu ions.}
\end{figure*}

Figure~\ref{Fig-cu-xmcd} summarizes the XAS data of the samples YL\_1-3 which reveal that the strength of the AF-EI between the Cu and Mn moments can be strongly suppressed whereas the magnitude of the Cu ions is hardly affected. Figure~\ref{Fig-cu-xmcd}~(a) displays a sketch of a YBCO/LCMO ML and the XMCD experiment which selectively probes the ordered moment of the Cu or Mn ions (along the x-ray propagation vector). The XMCD signal, i.e., the difference between the absorptions for the left and right circularly polarized x-rays, has been obtained in the total electron yield (TEY) and total fluorescence yield (TFY) modes. Due to the limited escape depth of the photo-electrons of only few nanometers, the TEY mode is very surface sensitive. Since the MLs are terminated with LCMO, the TEY Cu XMCD signal is governed by the Cu ions at the uppermost YBCO/LCMO interface. The TFY mode has a much larger probe depth and thus is equally sensitive to the bulk-like Cu ions away from the interface. {\par}
Figure~\ref{Fig-cu-xmcd}~(b) displays the XMCD at the Cu $L_{3,2}$-edges in the TEY mode for a low field of 0.5\,T and high fields of 5 or 6\,T. Figure~\ref{Fig-cu-xmcd}~(c) shows the corresponding magnetic field scans for the XCMD signals at the Cu and Mn edges. They have been taken at the energies where the XAS curves in Figure~\ref{Fig-cu-xmcd}~(b) exhibit the maximal XMCD signal. The positive (negative) XMCD (for the applied field, $H_{ext}{\textgreater}0$) indicates a magnetic moment that is parallel (antiparallel) to $H_{ext}$. For {\YLa}, the Cu XMCD is always negative and almost saturates above 1\,T. The corresponding Mn XMCD is positive with a similar saturation behavior. This is the signature of the antiparallel orientation of the Cu and Mn moments that was interpreted in terms of a strong AF-EI between Cu and Mn~\cite{Chakhalian2006}. {\par}
For {\YLb}, the Cu XMCD exhibits a remarkably different behavior. It is also negative at first, but above 1\,T it reveals a paramagnet-like trend with a zero crossing around 3\,T. The sign change is also evident from the Cu XMCD curves at 0.5\,T and 6\,T in Figure~\ref{Fig-cu-xmcd}~(b) (middle panel). Note that these trends have been confirmed for three different {\YLb} type samples. The observed behavior is characteristic of a very weak AF-EI between Cu and Mn that is eventually overcome by the Zeeman-interaction due to $H_{ext}$. Strikingly, similar XMCD field scans have indeed been reported for a molecular system with a weak AF coupling between Cr and Dy moments~\cite{Dreiser2012}. {\par}
Finally, for {\YLc}, the Cu XMCD signal is always positive suggesting that the AF-EI with Mn is entirely suppressed. This suppression of the AF-EI is expected since the Mn moments themselves are hardly ferromagnetic. The surprising result is that there is still no sign of a reduction of the magnitude of the maximal Cu XMCD signal and thus of the interfacial Cu moment. For the three kinds of samples YL\_1-3, the sum rule~\cite{Thole1992,Chen1995} analysis yields rather similar effective spin moments of $0.10{-}0.25$\,{\mB} per interfacial Cu ion.
\subsection{Two component scenario with paramagnetic bulk like Cu ions}\label{two-component}
First of all, it is important to clarify whether both the negative and the positive Cu XCMD signals originate from the CuO$_2$ planes next to the interface with LCMO or whether the latter is due to the bulk-like Cu ions in the more distant CuO$_2$ planes. The second possibility that the paramagnetic Cu XCMD originates from bulk-like Cu ions, has been suggested in Ref.~\cite{DeLuca2014} which reported similar XMCD data for corresponding LSCO/LCMO MLs. Nevertheless, as shown in the following, we can exclude such a two component scenario for the case of these YBCO/LCMO MLs. The evidence is obtained from the analysis of the multi-peak structure of the XAS curves (a detailed description can be found in Appendix~\ref{app-multipeak}). As shown in Figure~\ref{Fig-cu-xmcd}~(d)-(g), the interfacial and bulk-like Cu ions give rise to distinct peaks ~\cite{Uribe2014} in the XAS curves. In particular, the low-energy peak around 930.4\,eV (for which the position is marked by a dotted line in Figure~\ref{Fig-cu-xmcd}~(d)-(g)) originates from the interfacial Cu ions since it is much stronger in the TEY mode (Figure~\ref{Fig-cu-xmcd}~(e)) which primarily probes the Cu ions at the topmost YBCO/LCMO interface, than in the TFY mode (Figure~\ref{Fig-cu-xmcd}~(d)). The negative XLD ($\mu_{ab}-\mu_c$) of the $930.4$~eV peak, as compared to the positive one of the peaks at 931.0 and 931.7\,eV due to the bulk-like CuO$_2$ planes, is the fingerprint of the orbital reconstruction of the interfacial CuO$_2$ plane~\cite{Chakhalian2007,Uribe2014}. The red-shift of the peak at 930.4\,eV can be understood in terms of a charge transfer between LCMO and YBCO~\cite{Chien2013}. The Cu XMCD curves of {\YLb} in Figure~\ref{Fig-cu-xmcd}~(f) and~\ref{Fig-cu-xmcd}~(g) highlight that the XMCD originates predominantly from the peak at 930.4\,eV and thus from the interfacial Cu ions. This applies to the positive Cu XMCD signal at 6\,T as much as to the negative one at 0.5\,T. In Figure~\ref{Fig-cu-xmcd}~(g) there is also a very weak paramagnetic contribution from the peak at 931\,eV due to the bulk-like Cu ions. This paramagnetic signal is however at least an order of magnitude smaller than the one from the interfacial Cu ions. Note that this conclusion is independent of the particular fitting procedure that is used to account for the different peaks. The essential effect is even seen in the bare spectra where the position of the peak of $\mu_c$ in TEY mode (solid blue circles in Figure~\ref{Fig-cu-xmcd}~(e)) coincides with the one of the maximum in the XMCD signal (solid gray symbols in Figure~\ref{Fig-cu-xmcd}~(f)~and~(g)). An additional argument against the two-component scenario of Ref.~\cite{DeLuca2014} is that such a paramagnetic signal is not observed in {\YLa}, despite the circumstance that this sample contains the same amount of bulk-like Cu ions as {\YLb}.
\subsection{Depleted layer and the suppression of the FM order of interfacial Mn moments}\label{result-pnr}
\begin{figure}[!htbp] 
\centering
\includegraphics{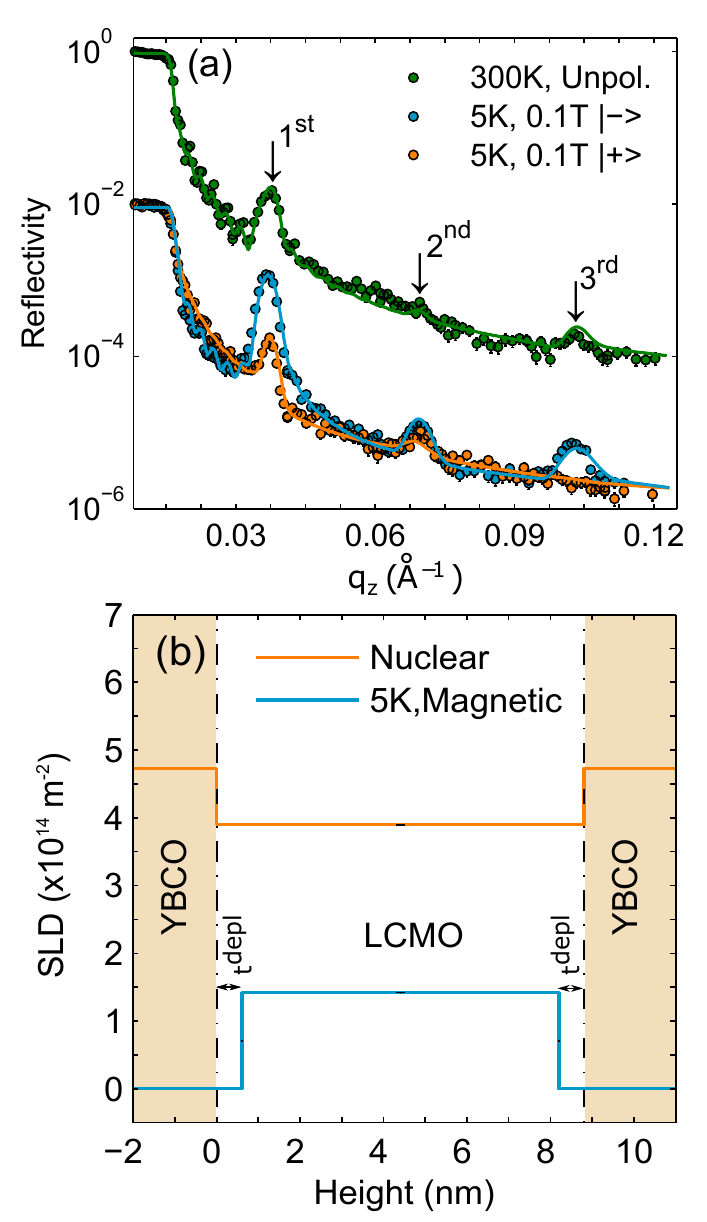}
\caption{\label{Fig-pnr-yl2} (a) Polarized neutron reflectometry data (symbols) for {\YLb} in the non-magnetic state at $300$\,K and the FM state at $5$\,K. Also shown are the best fits (solid lines) using a block-like depth profile of the nuclear and magnetic potentials. (b) The deduced nuclear (orange) and magnetic (blue) depth profiles in units of the scattering length density (SLD). The magnetic moment in the central part of LCMO amounts to about $3$\,{\mB} per Mn ion.}
\end{figure}

This puts the focus on the magnetic properties of the interfacial Cu ions and the questions, firstly, why the AF-EI with the Mn moments on the LCMO side is so strongly reduced in the {\YLb} type samples and, secondly, why the magnitude of the ferromagnetic Cu moment is independent of the strength of the exchange coupling with Mn. {\par}
Concerning the first question we note that, based on the x-ray diffraction and reflectometry data, the structural quality, and the roughness of the YBCO/LCMO interfaces are comparable for the {\YLa} and {\YLb} (10~BL repetitions) type samples. Furthermore, we find that whereas the roughness of the respective interface of the {\YLb} type sample with 1~BL repetition is only about 6\,{\AA} (discussed in Figure~\ref{fig-xrd}~(b)), the characteristics of the Cu XMCD signal is very similar to the one shown (in Figure~\ref{Fig-cu-xmcd}) for the {\YLb} 10~BLs sample that has a larger roughness of 10\,{\AA}. Therefore, the strong suppression of the AF-EI in {\YLb} is not the result of an increased interface roughness. Furthermore, the TEM study in Figure~\ref{fig-tem-eels} has shown that {\YLb} has the same interfacial stacking (termination) as it was reported in Ref.~\cite{Malik2012} for {\YLa}, i.e., with a sequence of CuO$_2$-BaO-MnO$_2$ layers that results in a straight Cu-O$_{apical}$-Mn bond across the YBCO/LCMO interface (and vice versa for the LCMO/YBCO interface). This kind of interface termination is corroborated by the Cu XLD data which show no significant differences between these samples (see Figure~1~(a)-(d) of Ref.~\cite{Uribe2014} for {\YLa}, Figure~\ref{Fig-cu-xmcd}~(d)~and~(e) for {\YLb}, and Figure~\ref{Cu-XLD-YL3}~(a)~and~(b) for {\YLc}). They suggest that these samples have a very similar charge transfer and orbital reconstruction of the interfacial Cu ions. In particular, the orbital reconstruction of the interfacial Cu ions, which is determined by the covalent bonding between the Cu and Mn ions, should be strongly affected by a change of the interfacial layer stacking. It was previously discussed that a strong covalent bonding  requires a  direct Cu-O$_{apical}$-Mn bond~\cite{Chakhalian2007} and shown that other kinds of interfacial layer stacking result in a much weaker orbital reconstruction effect~\cite{Uribe2014}.{\par}
Next, one may suspect that in {\YLb} the FM order of the Mn moments is more strongly suppressed at the interface with YBCO than in {\YLa}. This may be a result of the different growth and annealing conditions of the LCMO layers which in {\YLb} may weaken the FM double exchange interaction between the Mn moments and strengthen the competing AF interactions. Such a reduction of the Mn moment in the vicinity of the interfaces with YBCO has already been observed in the {\YLa} sample and was discussed in terms of a \textit{\lq{depleted layer}\rq}~\cite{Satapathy2012,Uribe2013}. Naturally, one might assume that this suppression of the FM order of the interfacial Mn moments is even stronger in {\YLb}. However, the PNR study of the same {\YLb} sample for which the XMCD data are shown in Fig~\ref{Fig-cu-xmcd} reveals the opposite trend, i.e., it shows that the thickness of the \textit{\lq{depleted layer}\rq} in {\YLb} is smaller than the one in {\YLa}. To enable a direct comparison, the PNR data of {\YLb} in Figure~\ref{Fig-pnr-yl2} have been fitted using the same block-like profile of the magnetic potential that was used in Ref.~\cite{Satapathy2012,Uribe2013} for {\YLa}. This simplistic model assumes a complete suppression of the FM order of the Mn moment in the \textit{\lq{depleted layer}\rq}. Therefore, it does not allow for the more realistic scenario of a gradual decrease of the FM moment and a finite value of the Mn moment at the interface. A stronger suppression of the interfacial Mn moment is thus accounted for in terms of a larger thickness of the \textit{\lq{depleted layer}\rq}, $t^{depl}$. The fitting shown in Figure~\ref{Fig-pnr-yl2}~(a) yields \textit{t$^{depl}$}$\approx{0.6}$~nm in {\YLb} that needs to be compared to \textit{t$^{depl}$}$\approx{1.5}$~nm in {\YLa} (the value is taken at the upper YBCO/LCMO interface~\cite{Satapathy2012,Uribe2013} that is probed in TEY mode). This trend has been confirmed with PNR measurements for an additional set of {\YLa} and {\YLb} type samples. This shows that a suppression of the FM order of the Mn moment near the interface can not explain the much weaker AF-EI in {\YLb}.
\subsection{Reduced AF-EI between Cu and Mn due to a change of the electronic/orbital properties of LCMO}\label{result-optics}
\begin{figure}[!htbp] 
\centering
\includegraphics{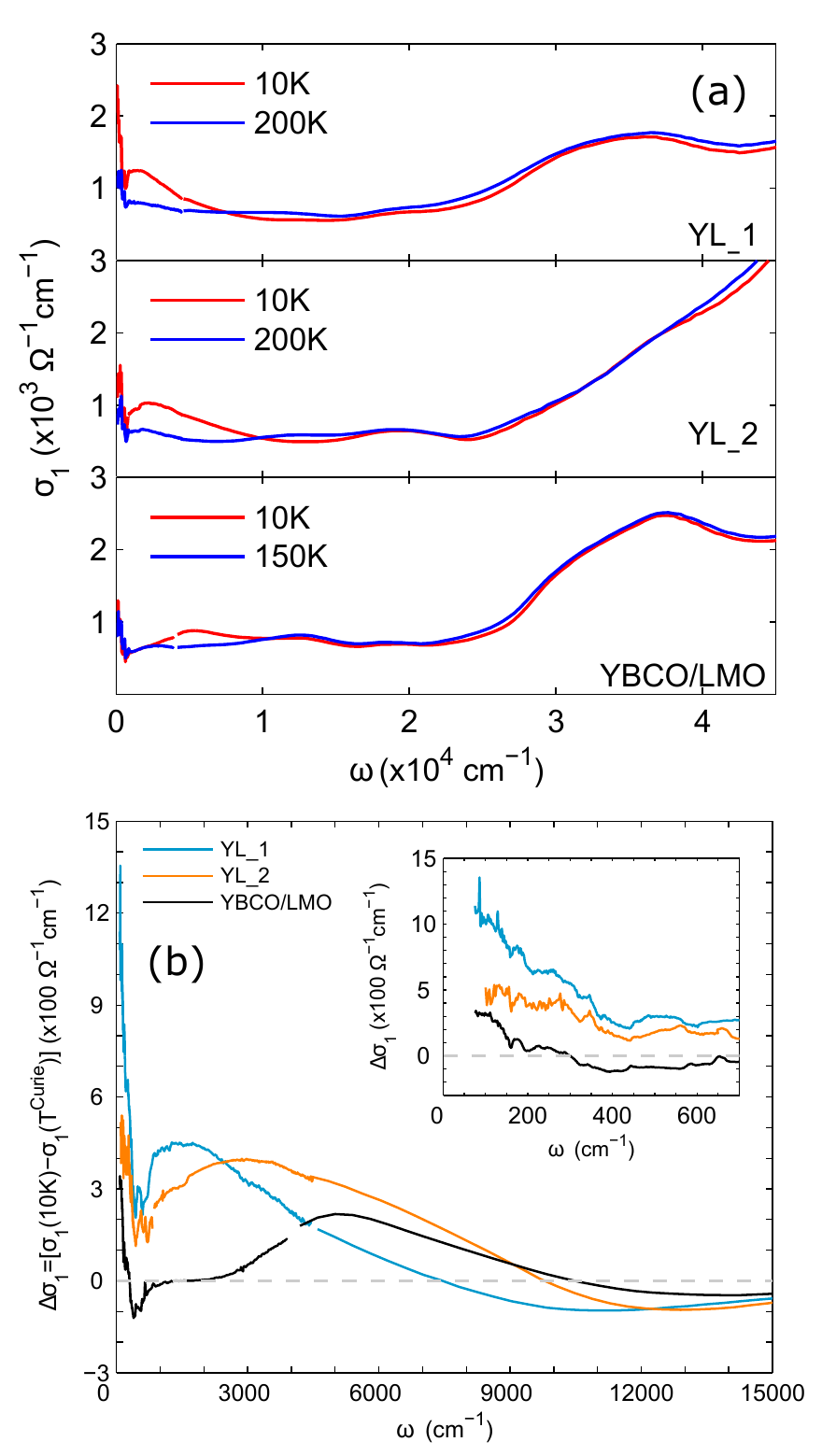}
\caption{\label{Fig-optics} (a) Real part of the optical conductivity spectra (${\sigma}_1$) of {\YLa}, {\YLb} and a [YBCO(10nm)/LaMnO$_{3+\delta}$(10nm)]$_{10}$ multilayer for which LMO is in the insulating FM state~\cite{Golod2013}. (b) Corresponding difference spectra of the optical conductivity, $\Delta\sigma_{1}(\omega)=\sigma_{1}(\omega,10K)-\sigma_{1}(\omega,T{\approx}T^{Curie})$. $\Delta\sigma_{1}$ of YBCO/LMO ML reveals a pronounced MIR band with a maximum around $5000$\,cm$^{-1}$  that is characteristic of the orbital polarons. The inset magnifies the low energy part of $\Delta\sigma_{1}$.}
\end{figure}

\begin{figure*}[!htbp] 
\centering
\includegraphics{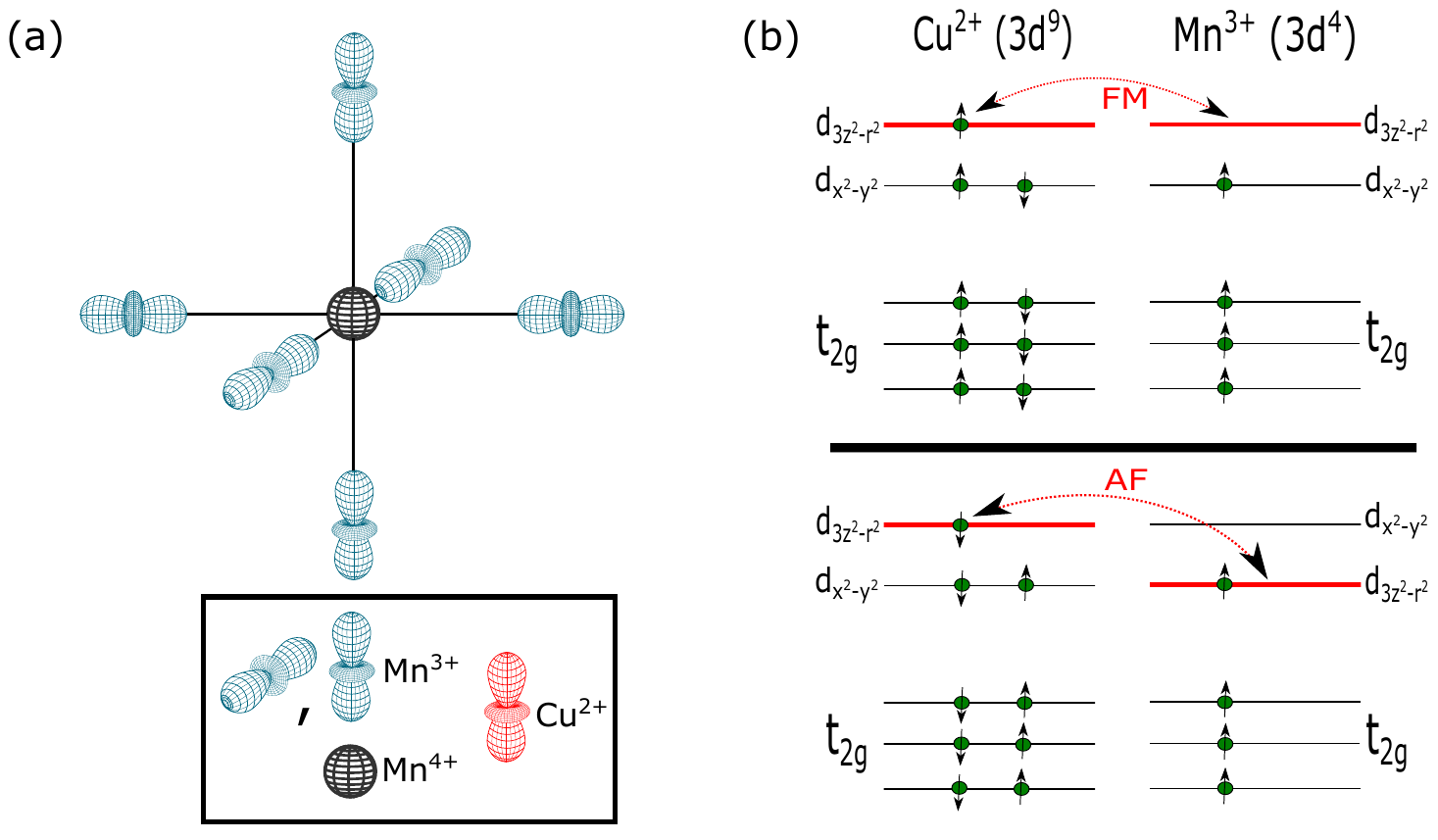}
\caption{\label{Fig7-polaron-coupling} (a) Sketch of an orbital polaron showing the occupied $e_g$ orbitals ($\ket{3z^2-r^2}$, $\ket{3x^2-r^2}$, and $\ket{3y^2-r^2}$). (b) Level scheme for interfacial Cu-$3d$ and Mn-$3d$ showing the change of the magnetic exchange interaction from FM (upper panel) to AF (lower panel) as the polarization of the occupied Mn-$e_g$ orbitals changes from the in-plane to the out-of-plane one. Shown is the idealized case of a 100\,\% orbital polarization due to the orbital reconstruction on the Cu-side.}
\end{figure*}

This puts the emphasis on the exchange coupling mechanism across the interface. It was previously shown that the YBCO/LCMO MLs have an interface termination with a layer stacking sequence of CuO$_2$-Y-CuO$_2$-BaO-MnO$_2$-(La,Ca)O at both the YBCO/LCMO and LCMO/YBCO interfaces (one plane of CuO chains per YBCO layer is missing)~\cite{Varela2003,Malik2012}. This termination thus yields a covalent Cu-O-Mn bond across the interface which according to Ref.~\cite{Chakhalian2007} is at the heart of the charge transfer and the orbital reconstruction of the interfacial Cu ions. The corresponding feature in the Cu XAS spectra is the red-shifted peak at 930.4\,eV that is strongly enhanced in TEY mode and exhibits only a weak Cu XLD signal (as opposed to the large Cu XLD signal of the bulk-like Cu ions). We find that these characteristic features due to the charge transfer and the orbital reconstruction of the interfacial Cu ions are equally present in the Cu XLD spectra of the samples {\YLa} (see Figure~1~(a)-(d) of Ref.~\cite{Uribe2014}), {\YLb} (see Figure~\ref{Fig-cu-xmcd}~(d)~and~(e)) and {\YLc} (see Appendix~\ref{app-xld-yl3}). In return, this suggests that these samples do not exhibit any significant differences concerning the interface termination, the resulting charge transfer and the orbital reconstruction of the interfacial Cu ions.
In the following we show that the most significant changes occur indeed with respect to the electronic properties of the LCMO layers. A clear reduction of the conductivity of the LCMO layers in {\YLb}, as compared to the one in {\YLa}, is evident from electric transport data in Figure~\ref{fig-rt-rh} and also from the infrared spectroscopy data in Figure~\ref{Fig-optics}. The R-T data in Figure~\ref{fig-rt-rh}~(a) show that the kink in the resistance around {\tcurie}, which is a signature of the transition from a paramagnetic insulator to a ferromagnetic metal in the LCMO layers, is fairly pronounced for {\YLa} whereas it is hardly visible for {\YLb}. The R-H curves in Figure~\ref{fig-rt-rh}~(b) show that the corresponding magneto-resistance effect in the vicinity of {\tcurie} is much smaller for  {\YLb} than for {\YLa}. Finally, the infrared spectra in Figure~\ref{Fig-optics}~(b) confirm that the increase of the Drude peak below {\tcurie}, which is a hallmark of the concomitant insulator-to-metal and paramagnet-to-FM transition of LCMO~\cite{Okimoto1997}, is much weaker in {\YLb} than in {\YLa}. For the former a significant fraction of the spectral weight is instead accumulated in a broad mid-infrared (MIR) band. The development below {\tcurie} of such a MIR band is well known from bulk manganites that are in the less hole doped part of the phase diagram close to the insulating FM (I-FM) phase where the charge carriers start to form FM polarons~\cite{Okimoto1997}. The signatures of this MIR band are also seen in a corresponding ML with LaMnO$_{3+\delta}$ (LMO) layers that are known to be in the I-FM state~\cite{Golod2013,Marozau2014} (see Figure~\ref{Fig-optics}~(b)). This MIR band is therefore a fingerprint of the FM polarons which are predicted to give rise to a particular local orbital order~\cite{Kilian1999} which involves an alternating occupation of the $3d${-}$e_{g}$ orbitals of the Mn$^{3+}$ ions as shown in Figure~\ref{Fig7-polaron-coupling}~(a). {\par}
Notably, the orbital occupation can have a profound effect on the exchange interaction with the interfacial Cu ions. This is shown by the scheme in Figure~\ref{Fig7-polaron-coupling}~(b) which illustrates that the exchange interaction is AF if the Mn $3d_{3z^2-r^2}$ level is occupied, whereas it becomes FM if the in-plane polarized orbital is occupied. Shown, for simplicity, is the extreme case for which the orbital reconstruction yields a complete inversion of the occupation of the Cu $3d_{x^2-y^2}$ and $3d_{3z^2-r^2}$ orbitals.
\begin{figure*}[!htbp] 
\centering
\includegraphics{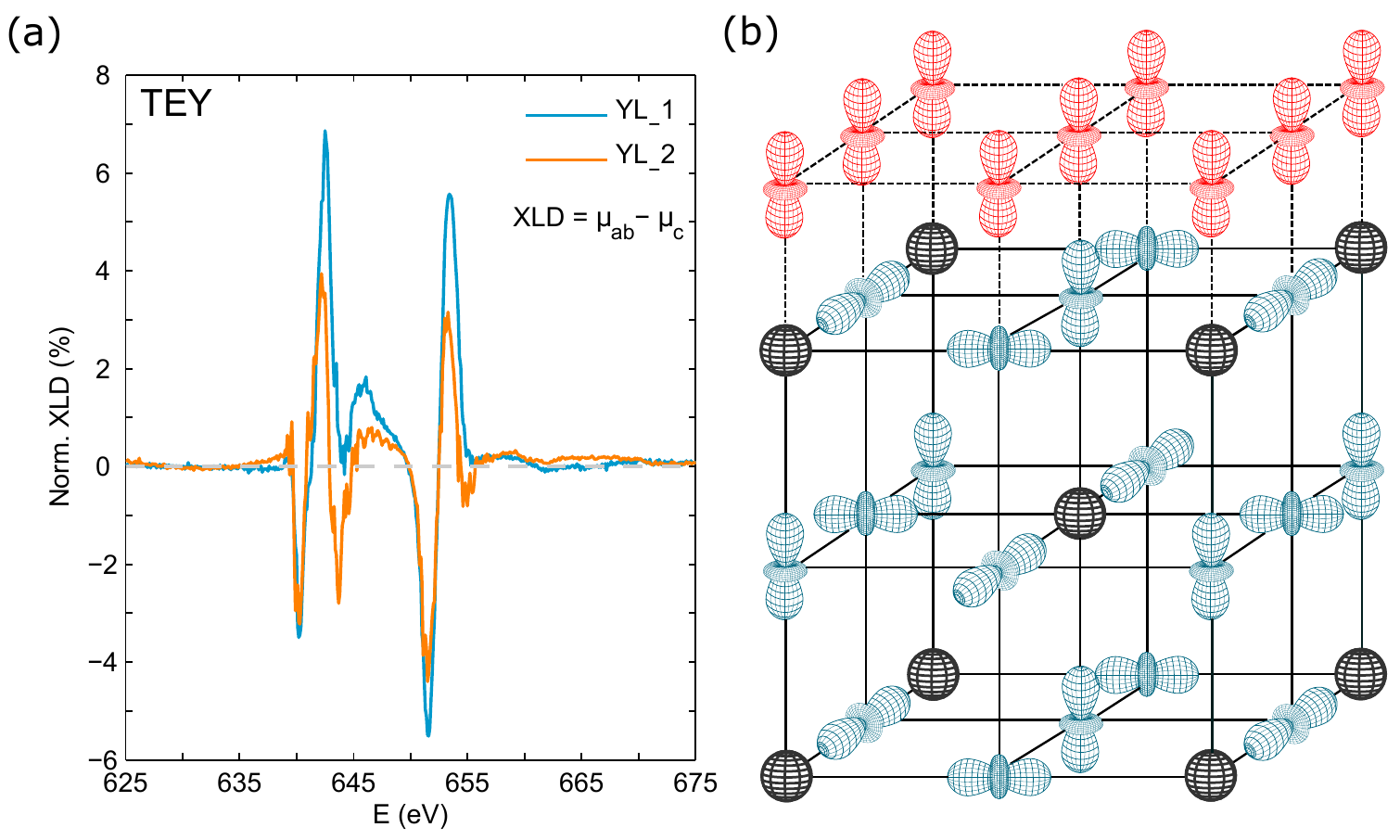}
\caption{\label{Figure8-MnXLD} (a) Mn XLD curves in TEY mode of {\YLa} and {\YLb} showing the larger orbital polarization in the former. (b) Sketch of the LCMO/YBCO interface for a lattice of the orbital polarons (in a FM state at 25\,\% hole doping). For YBCO only the $3d_{3z^2-r^2}$ orbitals are shown which participate in the exchange interaction.}
\end{figure*}
The strong AF-EI in {\YLa} accordingly can be understood in terms of a preferred occupation of the Mn $3d_{3z^2-r^2}$ orbitals. Such an effect is indeed suggested by the TEY Mn XLD data in Figure~\ref{Figure8-MnXLD}~(a) which yield an electron polarization of the $e_g$-orbitals of P$_{e_g}$$\approx+14$\,\% (as detailed in the Appendix~\ref{app-peg}). On the other hand, the scheme in Figure~\ref{Figure8-MnXLD}~(b) shows how the presence of the FM polarons in {\YLb} can reduce the strength of the AF-EI. It displays the spatial arrangement of the occupied Cu and Mn $e_g$-orbitals close to the interface due to a FM polaron lattice for the representative case of a doping of $x=0.25$. It gives rise to an alternation of the in-plane and out-of-plane polarization of the occupied Mn $e_g$-orbitals along the lateral direction and, accordingly (see Figure~\ref{Fig7-polaron-coupling}~(b)), to a sign change of the exchange interaction with the Cu ions. Needless to say that the scheme in Figure~\ref{Figure8-MnXLD}~(b) shows a simplified and qualitative picture of how the presence of the FM polarons leads to a reduction of the AF-EI between the interfacial Mn and Cu moments, even in the presence of a strong FM order of the Mn moments. In the LCMO layers of {\YLb}, these FM polarons may still be partially dynamic and strongly disordered. Nevertheless, especially in the vicinity of the interfaces, they can be pinned and thus contribute significantly to the suppression of the net AF-EI with the Cu moments. The presence of these FM polarons can also explain the smaller thickness of the \textit{\lq{depleted layer}\rq} in {\YLb} since they make the FM order of the Mn moments more robust against the interfacial strain and disorder effects. Finally, we remark that in line with this orbital polaron scenario, for which the average polarization of the Mn $e_g$-orbitals should vanish (see Appendix~\ref{app-polaron-xld}), the Mn XLD signal of {\YLb} in Figure~\ref{Figure8-MnXLD}~(a) yields a reduced value of P$_{e_g}$$\approx+3$\,\%.
\begin{figure}[!htbp]
\centering
\includegraphics{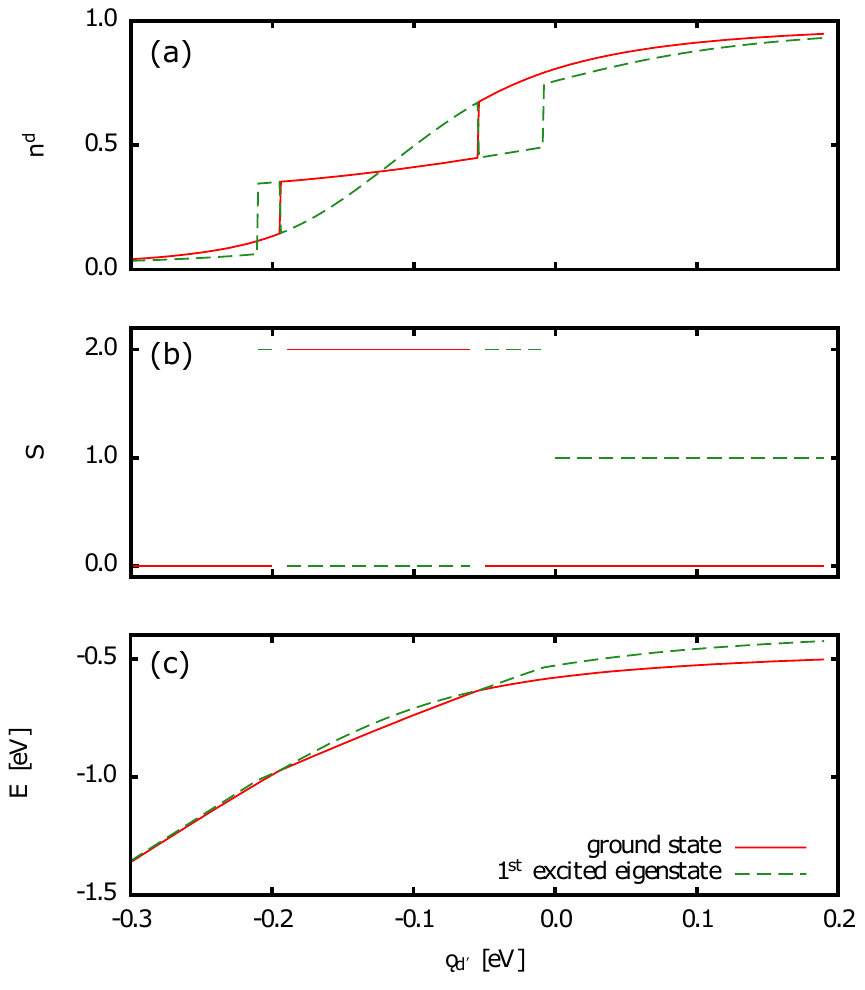}
\caption{\label{FIG9-10}
(a) The quantity $\langle n^{d}\rangle$---the number of holes in the orbitals $3d_{x^{2}-y^{2}}$ (per site)---as a function of the parameter $\epsilon_{d'}$. (b) $\epsilon_{d'}$ dependence of the total spin of the cluster. In both panels, the solid red line corresponds to the ground state and the green dashed line to the first excited state. (c) Energies of the ground state and of the first excited state as functions
of the parameter $\epsilon_{d'}$.}
\end{figure}
\begin{figure*}[!htbp]
\centering
\includegraphics{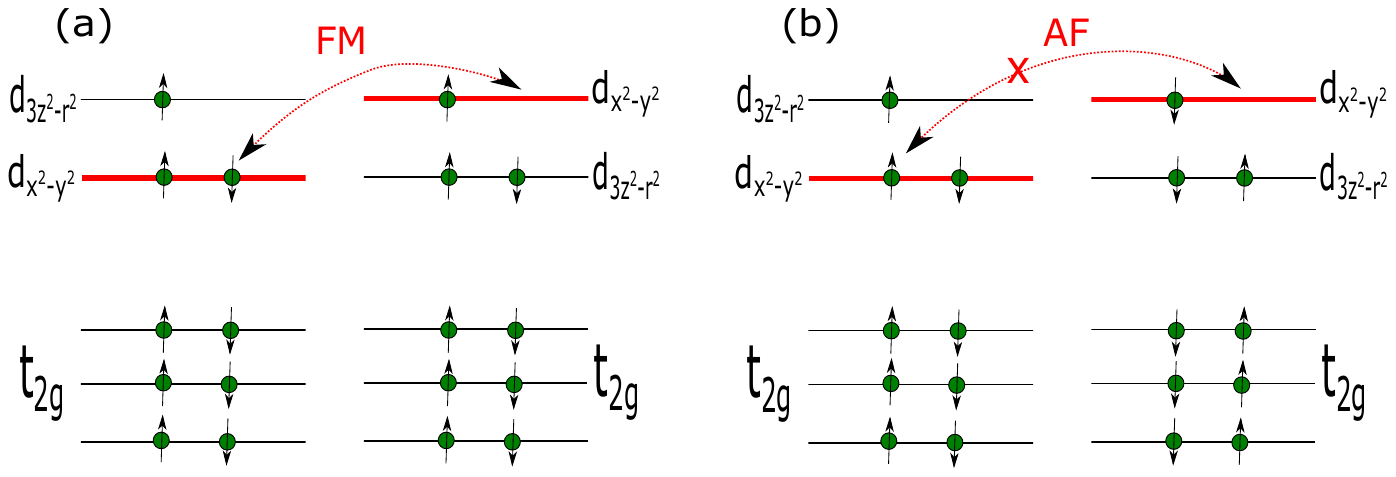}
\caption{\label{S4C} Schematic representation
of the virtual hoppings within the $d-d'$ configurations
discussed in the text, in the electron representation.
The hopping for the configuration of parallel spins indicated in (a)
results in an intermediate state of a lower energy
than the hopping for the configuration of antiparallel spins indicated in (b)
because of the Hund's rule coupling on the left site.}
\end{figure*}
\subsection{Intrinsic ferromagnetic order of the interfacial Cu ions}\label{result-yltheory}
This brings us to the second question, that has been formulated at the beginning of subsection~\ref{result-pnr}, about the origin of the intrinsic Cu moment of the interfacial CuO$_2$ plane. We suggest that this is due to its strongly underdoped state which arises from the electron transfer from LCMO to YBCO and the lack of a CuO chain layer which serves as a charge reservoir~\cite{Varela2003,Malik2012} and to the orbital reconstruction. In a bulk-like environment, where the holes reside mainly in Cu $3d_{x^2-y^2}$ orbitals, there would be a strongly AF intra-planar exchange interaction between the Cu moments which suppresses the corresponding Cu XMCD signal. Nevertheless, for the Cu ions in the CuO$_2$ plane next to the YBCO/LCMO interface, the occupation of the $e_g$-orbitals is substantially different: the concentration of holes in the Cu $3d_{3z^2-r^2}$ orbitals is approximately the same as in the Cu $3d_{x^2-y^2}$ orbtials. This is the so called orbital reconstruction effect that was reported in Ref.~\cite{Chakhalian2007} and is also clearly evident from the XLD curves in Figures~\ref{Fig-cu-xmcd}~(d)~and~(e). In the following we outline that this particular occupation of the Cu $3d$-$e_g$ orbitals weakens the intra-planar AF interaction and may even induce a weak FM interaction. {\par}
In order to obtain more insight into the origin of the magnetic moment and its relation to the orbital reconstruction, we have performed exact diagonalization calculations for a cluster containing four Cu sites with one hole per site, described by the extended Hubbard model~\cite{Oles1983,Oles2005}. The Hamiltonian in~the~hole representation reads
$$
H=\epsilon_d \sum_{\vi\sigma} n^d_{\vi\sigma}
+ \epsilon_{d^\prime} \sum_{\vi\sigma} n^{d^\prime}_{\vi\sigma}+
$$
$$
T_{dd}
\sum_{\langle \vi \vj \rangle \sigma}
\biggl( d^\dag_{\vi\sigma} d_{\vj\sigma} + H.c.\biggr)
+ T_{d^\prime d^\prime}
\sum_{\langle \vi \vj \rangle \sigma}
\biggl( d^{\prime\dag}_{\vi\sigma} d^\prime_{\vj\sigma} + H.c.\biggr)+
$$
$$
T_{dd^\prime}
\sum_{\langle \vi \vj \rangle \sigma}
\biggl(p_{\langle \vi \vj \rangle}\bigl[d^{\prime\dag}_{\vi\sigma} d_{\vj\sigma}+
d^\dag_{\vi\sigma} d^\prime_{\vj\sigma}\bigr] + H.c.\biggr)+
$$
$$
+ U \sum_{\vi\gamma=d,d^\prime} n^\gamma_{\vi\uparrow}n^\gamma_{\vi\downarrow}
+ U'\sum_{\vi\sigma\sigma'} n^d_{\vi\sigma} n_{\vi\sigma\prime}^{d'}+
$$
\begin{equation}
K \sum_{\vi\sigma\sigma'} d^\dag_{\vi\sigma} d'^\dag_{\vi\sigma'} d_{\vi\sigma'} d'_{\vi\sigma}
+  K \sum_\vi \biggl( d^\dag_{\vi\uparrow} d^\dag_{\vi\downarrow} d'_{\vi\downarrow}d'_{\vi\uparrow} + H.c. \biggr)\ .
\label{eq:Hubbard}
\end{equation}
It involves Cu $3d_{x^2-y^2}$ orbitals (operators $d$, $d^\dag$) and Cu $3d_{3z^2-r^2}$ orbitals ($d'$, $d'^\dag$),
the corresponding particle number operators are denoted by
$n^{d/d'}_{\vi\uparrow/\downarrow}$.
The first line of Eq.~(\ref{eq:Hubbard}) contains the on-site one-hole terms.
The symbols $\epsilon_{d}$ and $\epsilon_{d'}$ denote the energies
of the orbitals $3d_{x^2-y^2}$ and $3d_{3z^2-r^2}$, respectively.
Here, we set $\epsilon_{d}=0$.
The parameter $\epsilon_{d'}$ controls
the occupations of the orbitals:
for high (low) values of $\epsilon_{d'}$
the holes can be expected to reside mainly in the $3d_{x^{2}-y^{2}}$ orbitals (in the $3d_{3z^{2}-r^{2}}$ orbitals),
for intermediate values comparable occupations can be expected.
The second and the third lines of Eq.~(\ref{eq:Hubbard}) contain the hopping terms.
The sums run over all pairs of nearest neighbours
and $p_{\langle \vi \vj \rangle}$ is equal to 1 (-1)
for pairs oriented along the $y$ axis ($x$ axis).
Finally, the fourth and the fifth lines of Eq.~(\ref{eq:Hubbard}) contain the on-site interaction terms.
The third of them involves the Hund's rule exchange and can be written
as
\begin{equation}
-J_{\mathrm{Hund}}\sum_\vi \bigl[{\mathbf S}^{d}_{\vi}{\mathbf S}^{d'}_{\vi}+(1/4)n^{d}_{\vi}n^{d'}_{\vi}\bigr],
\label{eq:Hund}
\end{equation}
where $J_{\mathrm{Hund}}=2K$,
${\mathbf S}^{d/d'}_{i}$ are the on-site spin operators and
$n^{d/d'}_{\vi}=n^{d/d'}_{\vi\uparrow}+n^{d/d'}_{\vi\downarrow}$.
The calculations have been performed using the open boundary conditions.
The following values of the input parameters have been used:
$T_{dd}=0.35{\rm\,eV}$, $T_{dd^\prime}=T_{dd}/\sqrt{3}$, $T_{d^\prime d^\prime}=T_{dd}/3$,
$U=4{\rm\,eV}$, $K=0.6{\rm\,eV}$, $U'=U-2K$.
The dependence of the ground state properties on the remaining parameter $\epsilon_{d'}$ has been investigated. {\par}
Figure~\ref{FIG9-10}~(a) shows the expectation value $\langle n^{d}\rangle$
of the number of holes in the $3d_{x^{2}-y^{2}}$ orbitals (per site)
as a function of $\epsilon_{d'}$.
The solid red line corresponds to the ground state and
the dashed green line to the first excited state.
It can be seen, that for high (low) values of $\epsilon_{d'}$,
$\langle n^{d}\rangle$ of the ground state is close to 1 (close to 0), as expected.
For intermediate values in the range from ca -0.2 eV to ca -0.05 eV, $\langle n^{d}\rangle \approx 0.4$
and $\langle n^{d'}\rangle=1-\langle n^{d}\rangle \approx 0.6$.
Clearly, the ground state properties for this particular range
may be of relevance in the context of considerations
of orbitally reconstructed CuO$_{2}$ planes at the YBCO/LCMO interfaces. Figure~\ref{FIG9-10}~(b) shows the $\epsilon_{d'}$ dependence of the total spin of the cluster
in its ground state and in the first excited state.
Interestingly, with decreasing $\epsilon_{d'}$ the spin state of the ground state changes
from a singlet at high values of $\epsilon_{d'}$
to a quintuplet (spins of the four holes parallel)
to a singlet at low values of $\epsilon_{d'}$.
The quintuplet occurs precisely in the same $\epsilon_{d'}$-window
as the 0.4-plateau in $\langle n^{d}\rangle$.
For completeness we show in Figure~\ref{FIG9-10}~(c) the $\epsilon_{d'}$ dependence of the energies
of the ground state and of the first excited state.
Note the crossings between the singlet line and the quintuplet line. {\par}
The trend of the total spin can be qualitatively understood as follows.
For high values of $\epsilon_{d'}$ the holes are located mainly
in the $3d_{x^{2}-y^{2}}$ orbitals and the standard superexchange mechanism
involving virtual hoppings between these orbitals
stabilizes the singlet ground state.
In the large $U$ limit the ground state can be even described analytically~\cite{Fazekas1999}.
With decreasing  $\epsilon_{d'}$, more and more holes enter the $3d_{3z^{2}-r^{2}}$ orbitals,
and at certain critical value of $\epsilon_{d'}$,
virtual hoppings within $d-d'$ configurations begin to play a more important role than those within the $d-d$ configurations
and stabilize the ferromagnetic (quintuplet) ground state.
The reason for why the former hoppings support the ferromagnetic configuration
is that for parallel spins the energies of the intermediate states are lower than for antiparallel spins
because of the Hund's rule coupling (for a schematic representation of the mechanism, see Figure~\ref{S4C}). {\par}
The {\lq{ferromagnetic window}\rq}  in the $\epsilon_{d'}$ dependence of $S$
appears only for relatively high values of $J_{\mathrm{Hund}}$,
for the present values of $T_{dd}$ and $U$, $J_{\mathrm{Hund}}$ has to be larger than $0.96{\rm\,eV}$.
For very low values of $\epsilon_{d'}$, the holes are located mainly in the $3d_{3z^{2}-r^{2}}$ orbitals
and the superexchange mechanism involving virtual $d'-d'$ hoppings
yields the singlet ground state. {\par}
In conclusion, results of our calculations demonstrate that the orbital reconstruction
can give rise to a weak intra-planar ferromagnetic interaction between the spins of the interfacial Cu ions.
Note that the 3D versions of the $e_{g}$ Hubbard model are known to exhibit ferromagnetic solutions
for certain ranges of doping,
in particular for 1/4 filling and 3/4 filling, see, e.g., Ref.~\cite{Peters2010}.
\section{Summary}
We have shown, based on x-ray absorption spectroscopy and polarized neutron reflectometry measurements, that the antiferromagnetic exchange interaction (AF-EI) between
the interfacial Cu and Mn moments in YBCO/LCMO multilayers can be strongly suppressed, whereas the ferromagnetic moment of the interfacial Cu ions remains sizeable. This suggests that the Cu moments are not induced by the AF-EI with Mn but are intrinsic to the interfacial CuO$_2$ planes. We have outlined that a weakly ferromagnetic intra-planar magnetic exchange interaction between the Cu moments may arise due to the nearly equal hole occupation of the Cu $3d_{3z^2-r^2}$ and the $3d_{x^2-y^2}$ orbitals that is brought about by the so-called orbital reconstruction which originates from the hybridization with the Mn ions. We have furthermore suggested that the strong suppression of the AF-EI between the Cu and Mn moments may be caused by ferromagnetic polarons which develop in poorly hole doped LCMO layers. {\par}
In terms of applications, these findings may be used to create spin-active cuprate/manganite interfaces for which the relative orientation of the Cu and Mn moments can be readily varied with an external magnetic field. This kind of interfacial spin control can for example allow one to induce a spin-triplet superconducting order parameter which serves as source of spin-polarized supercurrents. The fabrication of mesoscopic devices in which these ideas can be tested remains a project for future research.
\begin{acknowledgments}
The work at UniFr has been supported by the Schweizer Nationalfonds (SNF) through grants No. 200020-153660 and CRSII2-154410/1. The work at Muni was supported by the projects CEITEC(CZ.1.05/1.1.00/02.0068) and MUNI/A/1496/2014. The XAS measurements have been performed at the EPFL/PSI XTreme beamline of the Swiss Light Source at the Paul Scherrer Institut, Villigen, Switzerland. The neutron experiment has been done at the NREX instrument operated by the Max-Planck Society at the Heinz Maier-Leibnitz Zentrum (MLZ), Garching, Germany and was supported by the European Commission under the $7^{th}$ Framework Programme through the ``Research Infrastructures" action of the Capacities Programme, NMI3-II, Grant Agreement number $283883$.
\end{acknowledgments}

\begin{appendix}
\section{Analysis of the substructure of the XAS curves}{\label{app-multipeak}}
The method of the multi-peak fitting of the XAS curves in TEY and TFY modes near their ${L_3}$-edge is discussed here. We have followed the same approach as detailed in Ref.~\cite{Uribe2014}. For the Cu atoms, we consider four transitions. These are (a) and (b) the $\mathrm{2p^63d^9 \rightarrow 2p^53d^{10}}$ transitions of the interfacial and the bulk $\text{Cu}^{2+}$ ions, respectively, and (c) and (d) the $\mathrm{2p^63d^9\underline{L} \rightarrow 2p^53d^{10}\underline{L}}$ transitions related to the Zhang-Rice singlets of bulk Cu ions in the $\text{CuO}_2$ planes and CuO chains, respectively. Accordingly, we used four Lorentzian functions for the fitting of the XAS near the ${L_3}$-edges [Eq.~\eqref{lorentzian}]. To account for the edge-jumps and the non-zero background under the XAS curves, we added a combination of linear and sigmodal functions [Eq.~\eqref{bg-multiplet}]. To find the peak positions, all TFY and TEY XAS curves of a given sample have been fitted simultaneously with the peak positions as common parameters. In a next step, the peak positions have been fixed and each pair of XAS curves ($\mu_{\mathrm{+}}$ and $\mu_{\mathrm{-}}$; $\mu_{ab}$ and $\mu_{c}$) are fitted simultaneously  in TFY and TEY modes by considering the parameters for background and widths of the Lorentzian profiles as common parameters. In this way, we determined the four peaks around $930.4$, $931.0$, $931.7$ and $932.6$\,eV. The peak around $930.4$\,eV arises from the interfacial Cu ions since its weight is small in the TFY mode but very large in the TEY mode.
\begin{equation}\label{lorentzian}
y_{L}(E)={\scriptstyle \sum}_{i=1}^{4}\frac{2A_{i}}{\pi}\frac{w_{i}}{4(E-x_{c,i})^{2}+w_{i}^{2}}
\end{equation}
\begin{equation}\label{bg-multiplet}
y_{bg}(E)=y_{0}+y_{1}E+\frac{B}{1+exp(-k(E-x_{0}))}
\end{equation}
\section{Cu XLD and XMCD spectra of {\YLc}}
\begin{figure}[!htbp] 
\centering
\includegraphics{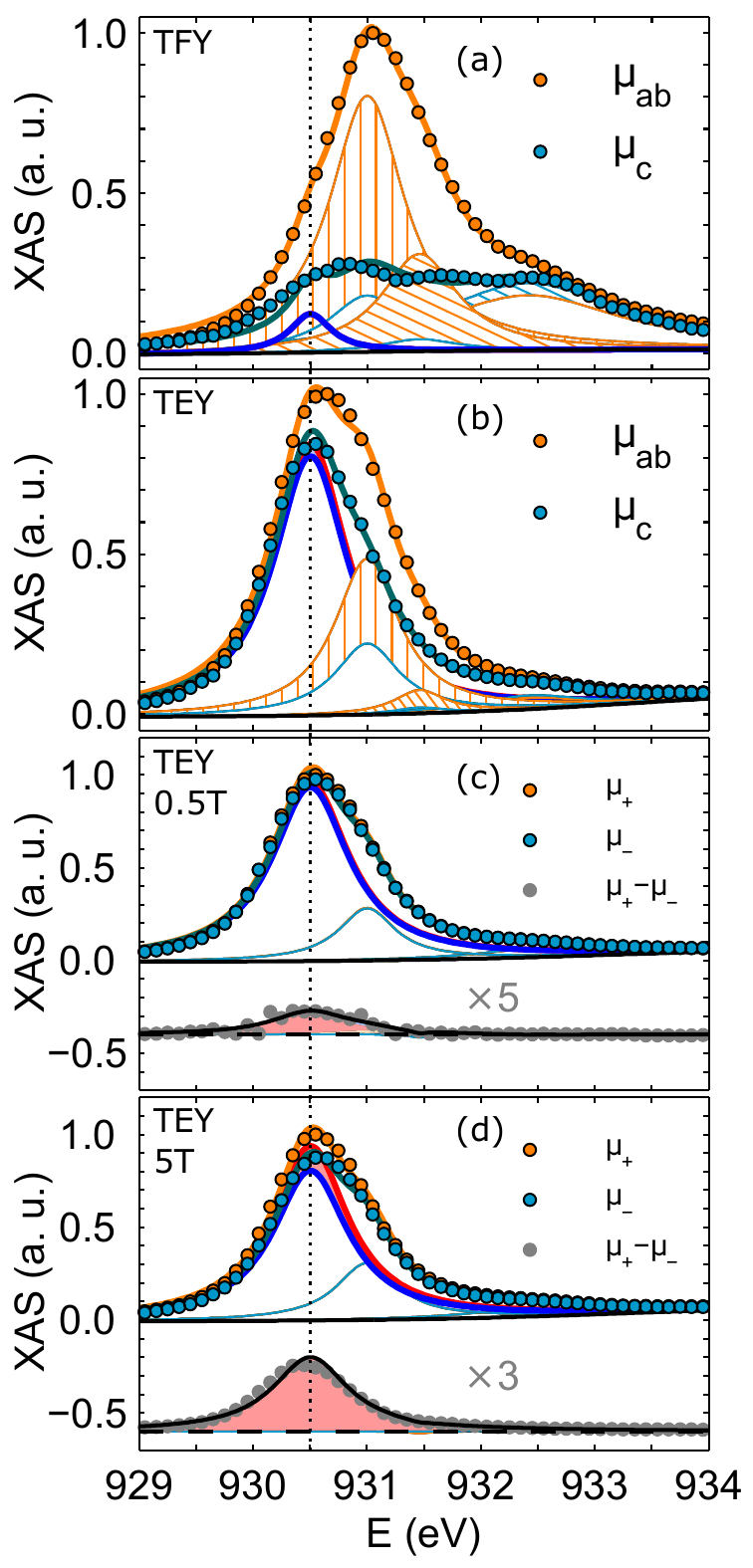}
\caption{\label{Cu-XLD-YL3} Cu XAS curves of {\YLc} with the linear polarization of the x-rays parallel and perpendicular to the CuO$_2$ planes in the (a) TFY and (b) TEY modes. Cu XAS curves of {\YLc} for circular polarizations and the corresponding XMCD signals at (c) 0.5\,T and (d) 5\,T.}
\end{figure}

{\label{app-xld-yl3}}
Similar to {\YLa} and {\YLb}, the signature of the orbital reconstruction of the interfacial Cu ions have been observed in {\YLc} as shown in Figures~\ref{Cu-XLD-YL3}~(a)~and~(b). The measurement has been done at 2\,K in remanence, after field cooling at 5\,T. The XAS curves in Figures~\ref{Cu-XLD-YL3}~(c)~and~(d) show the corresponding XMCD curves at 0.5 and 5\,T fields which confirm that the major part of the XMCD signal arises from the low energy peak at 930.5\,eV.
\section{Calculation of electron orbital polarization from Mn XLD data}
\begin{figure}[t] 
\centering
\includegraphics{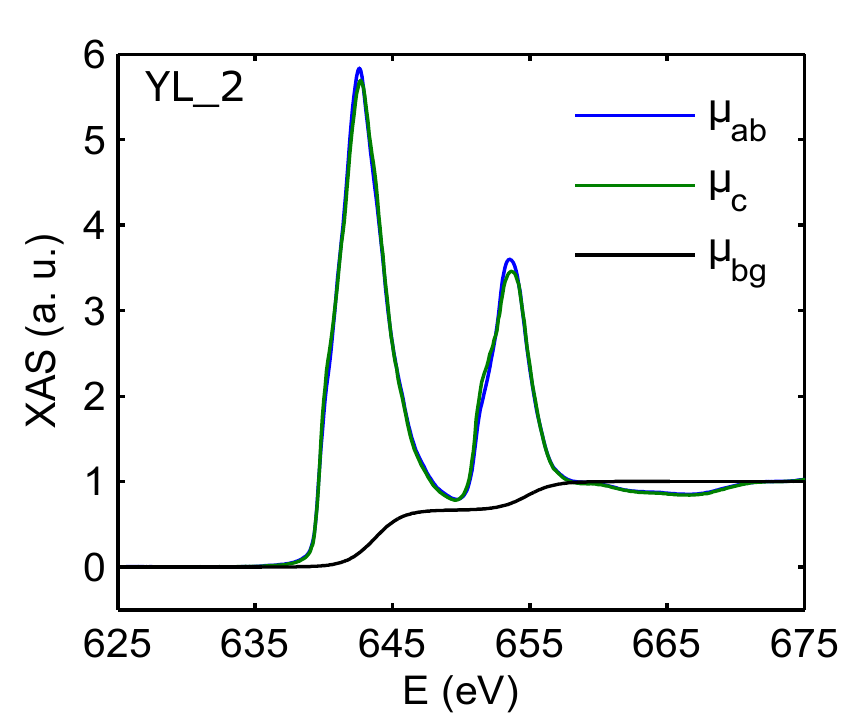}
\caption{\label{Mn-XLD-XAS} Mn XLD curves of {\YLb} in TEY mode showing $\mathrm{\mu_{ab}}$ and $\mathrm{\mu_{c}}$ and base line, $\mathrm{\mu_{bg}}$ which accounts for edge jumps.}
\end{figure}

{\label{app-peg}}
The electron orbital polarization, $\mathrm{P_{e_g}}$ of the LCMO layers has been deduced following the procedure of Refs.~\cite{Stohr1995,UribeThesis}.
\begin{equation}\label{peg}
\mathrm{P}_{e_{g}}=\frac{n_{3z^{2}-r^{2}}-n_{x^{2}-y^{2}}}{n_{3z^{2}-r^{2}}+n_{x^{2}-y^{2}}}=\frac{19}{2}\frac{\int_{L_{3}+L_{2}}2(\mu_{ab}-\mu_{c})d\omega}{\int_{L_{3}+L_{2}}(2\mu_{ab}+\mu_{c}-3\mu_{bg})d\omega}
\end{equation}
Here, ${n_{3z^2-r^2}}$ and ${n_{x^2-y^2}}$ are the number of electrons in the ${3d_{3z^2-r^2}}$ and ${3d_{x^2-y^2}}$ orbitals, respectively. Exemplary XAS curves for linear polarizations are shown in Figure~\ref{Mn-XLD-XAS}.
The positive sign of $\mathrm{P_{e_g}}$ indicates that ${3d_{x^2-y^2}}$  has a higher hole concentration, i.e., the electrons preferably occupy the ${3d_{3z^2-r^2}}$  orbital.
\section{Electron orbital polarization of $\mathbf{e_g}$-orbitals in a FM polaron}{\label{app-polaron-xld}}
It is shown that the ferromagnetic (FM) polarons do not give rise to x-ray linear dichroism signal (XLD), i.e., the elctron polarization of their $e_g$-orbitals amounts to zero. {\par}
As shown in Figure~\ref{Fig7-polaron-coupling}~(a), a FM polaron consists of two $3d_{3z^2-r^2}$, two $3d_{3x^2-r^2}$ and two $3d_{3y^2-r^2}$ orbitals. Among them, the last two orbitals are in-plane orbitals. In the following, they are represented in the basis of $\ket{x^2-y^2}$ and $\ket{3z^2-r^2}$. \\\\
$3d_{x^2-y^2}$\,:\,$\ket{x^2-y^2}$ \\
$3d_{3z^2-r^2}$\,:\,$\ket{3z^2-r^2}$  \\
$3d_{3x^2-r^2}$\,:\,$\frac{3}{2\sqrt{3}}\ket{x^2-y^2}-\frac{1}{2}\ket{3z^2-r^2}$ \\
$3d_{3y^2-r^2}$\,:\,$-\frac{3}{2\sqrt{3}}\ket{x^2-y^2}-\frac{1}{2}\ket{3z^2-r^2}$ \\\\
Calculation of the expectation value of the $\ket{x^2-y^2}$ and $\ket{3z^2-r^2}$ thus yields: \\\\
Number of $3d_{x^2-y^2}$ orbitals $=2{\times}0+2{\times}\frac{9}{12}+2{\times}\frac{9}{12}=3$. \\
Number of $3d_{3z2^2-r^2}$ orbitals $=2{\times}1+2{\times}\frac{1}{4}+2{\times}\frac{1}{4}=3$. \\\\
The $3d_{x^2-r^2}$ and $3d_{3z^2-r^2}$ orbitals thus have the same occupation probability which means that the XLD and the electron polarization of the $e_g$-orbitals amount to zero.
\end{appendix}

\end{document}